\documentclass[aps,letterpaper,prb,twocolumn,showpacs,superscriptaddress]{revtex4}
\usepackage{bm,color,amsmath,amssymb,graphicx}
\usepackage[pdftex,
            pdfauthor={Yang-Le Wu},
            pdftitle={}]{hyperref}

\providecommand{\Wsq}{W_{\protect\rule{3.5pt}{3.5pt}}}

\usepackage{array}
\newcolumntype{L}[1]{>{\raggedright\let\newline\\\arraybackslash\hspace{0pt}}m{#1}}
\newcolumntype{C}[1]{>{\centering\let\newline\\\arraybackslash\hspace{0pt}}m{#1}}
\newcolumntype{R}[1]{>{\raggedleft\let\newline\\\arraybackslash\hspace{0pt}}m{#1}}

\begin{document}

\title{Haldane Statistics for Fractional Chern Insulators with an Arbitrary 
Chern number}

\author{Yang-Le Wu}
\affiliation{Department of Physics, Princeton University, Princeton, NJ 08544}
\author{N. Regnault}
\affiliation{Department of Physics, Princeton University, Princeton, NJ 08544}
\affiliation{Laboratoire Pierre Aigrain, ENS and CNRS, 24 rue Lhomond, 75005 Paris, France}
\author{B. Andrei Bernevig}
\affiliation{Department of Physics, Princeton University, Princeton, NJ 08544}

\begin{abstract}
In this paper we provide analytical counting rules for the ground states and 
the quasiholes of fractional Chern insulators with an arbitrary Chern number.
We first construct pseudopotential Hamiltonians
for fractional Chern insulators.
We achieve this by mapping the lattice problem
to the lowest Landau level of a multicomponent
continuum quantum Hall system with specially engineered boundary conditions.
We then analyze the thin-torus limit of the pseudopotential Hamiltonians, and 
extract counting rules (generalized Pauli principles, or Haldane statistics) 
for the degeneracy of its zero modes in each Bloch momentum sector.
\end{abstract}
\maketitle

\section{Introduction}

As the canonical example of topological order, the fractional quantum Hall (FQH)
effect was originally discovered in two-dimensional electron gas subject to a 
strong perpendicular magnetic field.~\cite{Tsui82:FQH,Laughlin83:Nobel}
Recently, several groups demonstrated numerically that these
strongly-correlated phases 
also exist in a topological flat band characterized by a non-zero Chern number 
$C$, even in the absence of a magnetic 
field.~\cite{Sheng11:FCI,Neupert11:FCI,Regnault11:FCI}
This discovery of the so-called fractional Chern insulators (FCI) generated 
enormous interest.~\cite{Parameswaran13:Review,Bergholtz13:Review}
Subsequent numerical 
studies~\cite{Wang11:FCI-Boson,Neupert11:Z2,Bernevig12:Counting,Wu12:Zoology,
Wang12:MR,Venderbos12:t2g,Lauchli13:Hierarchy,Liu13:CF,Kourtis12:Triangular,
Kourtis13:Combined}
quickly confirmed the presence of more intricate single-component FQH states 
in lattice 
models~\cite{Haldane88:Honeycomb,Sun11:Flatband,Tang11:Kagome,Hu11:Ruby}, such as
the Read-Rezayi 
series~\cite{Moore91:MR,Read99:RR,Bernevig12:Counting,Wu12:Zoology,Wang12:MR}
and the composite-fermion states~\cite{Jain89:CF,Lauchli13:Hierarchy,Liu13:CF}.
Powerful techniques from the study of FQH, including 
density algebra,~\cite{Girvin86:GMP,Parameswaran12:W-inf,Bernevig12:Counting,
Goerbig12,Roy12:Geometry,Dobardzic13:GMP}
entanglement spectrum,~\cite{Li08:ES,Sterdyniak11:PES,Regnault11:FCI,Wu12:Zoology}
parton construction,~\cite{Wen99:Parton,Lu12:Parton,McGreevy12:Parton}
and the Hamiltonian theory of composite fermions,~\cite{Murthy11:CF,Murthy12:CF}
were introduced to understand the topological ground state of FCI 
and the nature of its excitations.~\cite{Wu12:Wannier,
Lee13:Wannier,Liu13:Wannier,Scaffidi12:Adiabatic,Wu12:Hofstadter,
Zhu13:MES,Liu13:Edge,Lee13:Pseudopotential}
Possible experimental realizations have also been 
proposed.~\cite{Cooper13:OFL,Yao13:Dipolar}

Most of the above progress dealt with a topological band with Chern number 
$C=1$, which is essentially the same~\cite{Haldane88:Honeycomb}
as the continuum FQH in a periodic
potential~\cite{Kol93:FQH-Lattice,Moller09:FQH-Lattice,Sorensen05:FQH-Lattice}.
The strongly-correlated physics in a $C>1$ Chern 
band~\cite{Wang11:Dice,Trescher12:Flatband,Yang12:Flatband,Wang12:C2,
Wu13:BlochHofstadter}
turned out much 
richer than the conventional FQH, due to the interplay between topological 
order and lattice structure.~\cite{Barkeshli12:Nematic,
Liu12:Higher,Sterdyniak13:Higher,Lu12:Parton,Barkeshli13:Twist}
Barkeshli and Qi~\cite{Barkeshli12:Nematic}
mapped a $C>1$ Chern band to a $C$-component lowest Landau 
level (LLL) using hybrid Wannier states~\cite{Qi11:Wavefunction},
and suggested the possibility to realize multicomponent FQH states in a single 
Chern band.
Numerical 
studies~\cite{Wang12:C2,Liu12:Higher,Sterdyniak13:Higher,Grushin12:Dispersion} 
indeed found clear signature of such states,
including the color $\mathrm{SU}(C)$ version of the 
Halperin~\cite{Halperin83} and the non-Abelian 
spin-singlet states~\cite{Ardonne99:Singlet} (NASS),
but also identified qualitative deviations from these 
states,~\cite{Sterdyniak13:Higher,Liu12:Higher}
which implies a more complex structure than proposed in 
Ref.~\onlinecite{Barkeshli12:Nematic}.
In a previous paper,~\cite{Wu13:Bloch}
we proposed to understand these new features as
the consequences of a special set of boundary conditions
associated with the LLL mapping.
In the simplest case, this alternative boundary condition can be understood as 
a color-dependent magnetic flux insertion.
We demonstrated that the multicomponent LLL in a new Bloch basis can be seen as
a single manifold with constant Berry curvature and Chern number $C$.
Using pseudopotential Hamiltonians, we constructed model states for FCI with 
an arbitrary Chern number,
and found high overlaps with the exact ground states.
Crucially, our model states correctly capture the anomalous features in the 
particle entanglement spectrum of the $C>1$ FCI that make our states distinct 
from the conventional multicomponent FQH states.

In this paper we provide details of the mapping between a Chern band and a 
multicomponent LLL, and demonstrate the distinctive features of our 
pseudopotential Hamiltonian due to the new boundary conditions.
We construct, in a $C$-component LLL, a momentum-space Bloch basis and a hybrid 
Wannier basis that mimic the lattice counterparts.
Both bases entangle the real space and the internal color space.
Using the explicit one-body wave functions for the bases, we derive the representation 
of the projected density operators in both bases.
We define model states as the exact zero modes of the pseudopotential 
Hamiltonian built from the projected density operators.
As we demonstrated in our previous paper~\cite{Wu13:Bloch},
the Bloch basis is useful for numerical 
studies as it preserves the full lattice symmetry.
The hybrid Wannier basis, on the other hand, facilitates the analysis of the 
pseudopotential Hamiltonian.

We give a detailed analysis of the simplest bosonic pseudopotential 
Hamiltonian for the Halperin color-entangled states.
We show that the pseudopotential Hamiltonian reduces to \emph{almost} 
classical electrostatics in the hybrid Wannier basis, when we take the 
so-called thin-torus 
limit~\cite{Tao83:TT,Bergholtz05:TT,Seidel06:TT,Bergholtz06:TT,Seidel07:TT,
Bergholtz08:TTCounting,Seidel08:mmn,Ardonne08:TT,Seidel08:MR,Bergholtz08:TT,
Seidel11:HR,Kardell11:TT,Nakamura12:LaughlinMPS,Bernevig12:TT}
and carry out truncations motivated by
previous numerical results.~\cite{Sterdyniak13:Higher,Liu12:Higher}
This enables us to write down the form of its zero modes in this limit.
However, in contrast to most well-known FQH states such as Laughlin and 
Read-Rezayi, a purely classical thin-torus description is not possible.
We pinpoint the key difference from the conventional multicomponent FQH due to 
a subtle twist in the hybrid Wannier states,
and detail the procedure to compute the total Bloch momentum of each zero mode.
The resulting algorithm correctly predicts the degeneracy of the FCI 
quasiholes in each lattice momentum sector, without resorting to numerical 
diagonalization, and can be seen as the extension of the generalized Pauli 
principle~\cite{Bernevig08:Jack,Bernevig08:Jack2} to the color-entangled states.

\section{One-Body States in a Multicomponent Lowest Landau Level}

In this section we construct one-body bases in a multicomponent LLL that mimic 
the Bloch and the Wannier bases in a Chern band with an arbitrary Chern number 
$C$.\footnote{In the following discussion we assume $C>0$ for simplicity. The 
case of $C<0$ can be handled by inverting, say, the $x$ direction of the Landau 
level.}
We consider a $C$-component (generalized spin)
electron moving on a torus with a perpendicular uniform magnetic field.
The major difference between our approach and the usual treatment of the 
multicomponent LLL problem is the adoption of a new set of boundary conditions.
This alternative choice entangles together the $C$ components and enables us to
construct a single manifold of Bloch states with Chern number $C$.
In contrast to the usual picture of multicomponent LLL as $C$ separate 
manifolds (one for each of the $C$ components) each with unity Chern number, our 
bases provide a natural foundation 
for the mapping to a single Chern band with an arbitrary Chern number $C$.
The central result of this Section is Eq.~\eqref{eq:density-Bloch}, the 
expansion of the electron density operator in the Bloch basis.

\subsection{Translations Operators}

We consider electrons with $C$ internal (color) degrees of freedom
\begin{equation}
|\sigma\rangle,\quad\sigma\in\mathbb{Z}_C.
\end{equation}
For simplicity, we work on a rectangular torus spanned by 
$\mathbf{L}_x=L_x\hat{x}$ and $\mathbf{L}_y=L_y\hat{y}$, where $L_x$ and 
$L_y$ are the two fundamental cycles of the torus, and $\hat{x}$ and 
$\hat{y}$ are orthonormal.
The torus is pierced by a magnetic field in the $-\hat{e}_z$ 
direction, $\mathbf{B}=\nabla\times\mathbf{A}=B\hat{e}_z$ with $B<0$.
We denote by $e<0$ the charge of the electron.
The magnetic length is $l_B=\sqrt{\hbar/(eB)}$. 
We define the total number of fluxes $N_\phi$ penetrating the 
torus by
\begin{equation}\label{eq:Nf-def}
L_xL_y=2\pi l_B^2 N_\phi.
\end{equation}
Here we do \emph{not} assume $N_\phi$ to be an integer as in the 
original treatment of the Landau level on a toroidal 
geometry~\cite{Haldane85:TorusBZ}.
As we will see soon, the alternative set of boundary conditions we pick only 
requires
\begin{equation}
C N_\phi\in\mathbb{Z}.
\end{equation}
This integer is equal to the dimension of the one-body Hilbert space in the lowest 
Landau level.
We define the magnetic translation operator
\begin{equation}
T(\mathbf{a})=e^{-i\mathbf{a}\cdot\mathbf{K}/\hbar},
\end{equation}
where 
\begin{equation}
\mathbf{K}=-i\hbar\nabla-e\mathbf{A}(\mathbf{r})
+e\mathbf{B}\times \mathbf{r}
\end{equation}
is the guiding center momentum.
The translation $T(\mathbf{a})$ commutes with the one-body Landau Hamiltonian
$H=(-i\hbar\nabla-e\mathbf{A})^2/(2m)$ but not with the translation 
$T(\mathbf{b})$ at a different displacement,
\begin{equation}
T(\mathbf{a})T(\mathbf{b})=T(\mathbf{b})T(\mathbf{a})
e^{i \hat{z}\cdot \mathbf{a}\times \mathbf{b}/l_B^2}.
\end{equation}

As argued in the introduction, we need to make contact between the 
multicomponent Landau level states and the Bloch states in a Chern band.
For the latter, we consider a single Bloch band with Chern number $C$
in a tight-binding model on a lattice with $N_x\times N_y$ unit cells.
The band has a total of $N_xN_y$ one-body states, one at each
lattice momentum in the $N_x\times N_y$ Brillouin zone (BZ).
To make contact with this lattice system,
we first look in the Landau level for a pair of commuting
translation operators that also resolve an $N_x\times N_y$ BZ.
To this end, we tune the magnetic field to match the number of one-body 
states,
\begin{equation}\label{eq:CNf-NxNy}
CN_\phi = N_xN_y,
\end{equation}
and we consider the magnetic translations over a \emph{fictitious} 
$N_x\times N_y$ unit cell structure of the continuous torus, namely,
\begin{equation}\label{eq:TxTy}
\begin{aligned}
T_x&=T(\mathbf{L}_x/N_x),&
T_y&=T(\mathbf{L}_y/N_y).
\end{aligned}
\end{equation}
The operator $T_x$ (resp. $T_y$) has $N_x$ (resp. $N_y$) different eigenvalues.
As opposed to the $C=1$ case, however, for generic $C$
they do not commute due to the $N_\phi/(N_xN_y)=1/C$ flux over 
each fictitious plaquette,
\begin{equation}
T_xT_y=T_yT_x e^{i2\pi/C}.
\end{equation}
To compensate for this, we define the `clock and shift' operators $Q$ and $P$ over 
the internal (color) Hilbert space by
\begin{align}\label{eq:clock-and-shift}
P|\sigma\rangle&=|\sigma+1\text{ (mod $C$)}\rangle,&
Q|\sigma\rangle&=e^{i2\pi\sigma/C}|\sigma\rangle.
\end{align}
Both operators are unitary, and they satisfy
\begin{equation}
PQ=QPe^{-i2\pi/C}.
\end{equation}
This leads to a pair of \emph{commuting} composite operators
\begin{equation}\label{eq:color-entangled-T}
\begin{aligned}
\widetilde{T}_x&=T_xP,&
\widetilde{T}_y&=T_yQ.
\end{aligned}
\end{equation}
We will refer to this pair as the `color-entangled' magnetic translation operators.
For the (color-neutral) Landau Hamiltonian, both
operators are good symmetries, and they resolve an $N_x\times N_y$ Brillouin 
zone once we specify the boundary conditions.
Notice that in general $[T(\mathbf{L}_x),\widetilde{T}_y]\neq 0$, 
$[T(\mathbf{L}_y),\widetilde{T}_x]\neq 0$.
This means that we have to abandon the usual boundaries~\cite{Haldane85:TorusBZ}
$T(\mathbf{L}_\alpha)=1$, $\alpha=x,y$.
Instead, we adopt the color-entangled generalization 
$\widetilde{T}_\alpha^{N_\alpha}=1$, namely,
\begin{equation}\label{eq:color-entangled-boundaries}
T(\mathbf{L}_x)P^{N_x}=T(\mathbf{L}_y)Q^{N_y}=1.
\end{equation}
This alternative set of boundary conditions make it possible to construct two 
sets of basis states in the one-body Hilbert space with desirable properties 
spelled below.

\subsection{Bloch and Wannier Bases}

We define the Bloch states $|\mathbf{k}\rangle$ as the simultaneous 
eigenstates of $\widetilde{T}_x$ and $\widetilde{T}_y$ within the LLL,
\begin{equation}\label{eq:Bloch-eigenvalue}
\widetilde{T}_\alpha|\mathbf{k}\rangle=
e^{-i2\pi k_\alpha/N_\alpha}|\mathbf{k}\rangle,
\end{equation}
with $\mathbf{k}=(k_x,k_y)\in\mathbb{Z}^2$.
The $N_xN_y$ states within the first Brillouin zone 
\begin{equation}
\mathrm{1BZ}=[0~..~N_x)\times [0~..~N_y)
\end{equation}
have distinct eigenvalues under 
$\widetilde{T}_\alpha$, and they constitute the Bloch basis in
the $N_xN_y=CN_\phi$-dimensional Hilbert space of the $C$-component LLL.

We now look for the explicit wave function 
$\langle x,y,\sigma|\mathbf{k}\rangle$ for these basis states.
We specialize to the Landau gauge $\mathbf{A}=Bx\hat{y}$.
Consider the states $|X,k_y\rangle$ with $X,k_y\in\mathbb{Z}$ defined by
the real- and internal-space wave function\footnote{Here we discuss
the hybrid Wannier states $|X,k_y\rangle$ for convenience.
We could also work with the alternative set of hybrid Wannier states
$|Y,k_x\rangle$ (localized in the $y$ direction).~\cite{Wu12:Wannier}
We do not assume anything special in $N_x$ vs. $N_y$.
In particular, we make no assumption in the commensuration
between $N_x$, $N_y$, and $C$.
}
\begin{multline}\label{eq:Wannier-wf}
\!\!\!\!
\langle x,y,\sigma|X,k_y\rangle
=\frac{1}{(\sqrt{\pi}L_yl_B)^{1/2}}\!\!
\sum_m^{\mathbb{Z}}
\delta_{\sigma,X+mN_x}^{\text{mod }C}\\
\exp\Big\{\textstyle
i2\pi\!\left(\frac{X N_y+k_y C}{C}+mN_\phi\right)\frac{y}{L_y}
\qquad\qquad\qquad\qquad\\
-\frac{1}{2}\!\left[
\frac{x}{l_B}-\frac{2\pi l_B}{L_y}\left(\frac{XN_y+k_y C}{C}+mN_\phi\right)
\right]^2
\Big\}.\!\!\!\!
\end{multline}
Here $X,k_y$ are state labels taking integer values, while $x,y$ 
are real space coordinates taking continuous values,
and $\sigma\in\mathbb{Z}_C$ is a discrete coordinate
in the internal color space.
It is not hard to see that $|X,k_y\rangle$ belongs to the lowest Landau level,
as the above wave function can be recast in the form 
$f(x+iy,\sigma)\,e^{-x^2/(2l_B^2)}$.
Moreover, we find that $|X,k_y\rangle$ is periodic in $X$, but with a twist in 
$k_y$:
\begin{equation}\label{eq:Wannier-periodicity}
\begin{aligned}
|X+N_x,k_y\rangle&=|X,k_y\rangle,\\
|X,k_y+N_y\rangle&=|X+C,k_y\rangle.
\end{aligned}
\end{equation}
These relations are reminiscent of the flow of hybrid Wannier states in a Chern 
insulator~\cite{Wu12:Wannier}. 
Moreover, as we prove in Appendix~\ref{sec:Wannier-translations},
the color-entangled magnetic translations [Eq.~\eqref{eq:color-entangled-T}] 
have a representation on 
$|X,k_y\rangle$ similar to the representation of the lattice translations on 
the hybrid Wannier states, namely,
\begin{equation}\label{eq:Wannier-translations}
\begin{aligned}
\widetilde{T}_x|X,k_y\rangle&=|X+1,k_y\rangle,\\
\widetilde{T}_y|X,k_y\rangle&=e^{-i2\pi k_y/N_y}|X,k_y\rangle.
\end{aligned}
\end{equation}
We thus refer to these states as the hybrid Wannier states
in the $C$-component LLL.
It is easy to see the states with $X\in[0~..~N_x)$ and $k_y\in[0~..~N_y)$ 
are linearly independent.
We emphasize that unless $N_x$ is divisible by $C$,
these states are \emph{not} color eigenstates,
in contrast to the states studied in Ref.~\onlinecite{Barkeshli12:Nematic}.

We want to define the Bloch states in the LLL as a Fourier sum of the hybrid 
Wannier states,\footnote{Here and hereafter, the summation of the shorthand 
form $\sum_a^{N}$ stands for $\sum_{a=0}^{N-1}$.}
\begin{equation}\label{eq:Bloch-def}
|\mathbf{k}\rangle=|k_x,k_y\rangle=\frac{1}{\sqrt{N_x}}\sum_{X}^{N_x}
e^{i2\pi Xk_x/N_x}|X,k_y\rangle.
\end{equation}
From Eqs.~\eqref{eq:Wannier-periodicity} and~\eqref{eq:Wannier-translations}, 
we find that the simultaneous eigenvalue equation 
in~\eqref{eq:Bloch-eigenvalue} indeed holds.
These states are periodic in $k_x$, but only quasi-periodic in $k_y$,
\begin{align}\label{eq:Bloch-periodicity}
|k_x+N_x,k_y\rangle&=|k_x,k_y\rangle,\\
|k_x,k_y+N_y\rangle&=e^{-i2\pi k_x C/N_x}|k_x,k_y\rangle.
\end{align}
The latter non-periodicity signals the topological obstruction to a periodic 
smooth gauge due to the non-zero Chern number of a Landau level.~\footnote{We 
can perform a gauge transformation to make the Bloch states periodic.
However, the resulting wave function will not be smooth in $k_x/N_x$ and/or 
$k_y/N_y$ in the continuum limit $N_x,N_y\rightarrow \infty$.
For example,
for $k_y\in[mN_y~..~mN_y+N_y)$ with $m\in\mathbb{Z}$, we can take
$|k_x,k_y\rangle\rightarrow e^{i2\pi k_x mC/N_x}|k_x,k_y\rangle$.
This transformation makes the state periodic, but discontinuous at 
$k_y/N_y\in\mathbb{Z}$.
}

\subsection{Projected Density Operator}

The density operator projected to the lowest Landau level plays a central role 
in the FQH physics, as it is used to define the inter-particle interaction.
As we now show, this operator takes a particularly nice form in our Bloch 
basis.

By definition, the density operator of color $\sigma$
at $\mathbf{r}=(x,y)$ projected to the LLL is given by
\begin{equation}
\rho(\mathbf{r},\sigma)
=\sum_{\mathbf{k}_1}^\mathrm{BZ}\sum_{\mathbf{k}_2}^\mathrm{BZ}
|\mathbf{k}_1\rangle
\phi^*_{\mathbf{k}_1}(\mathbf{r},\sigma)
\phi_{\mathbf{k}_2}(\mathbf{r},\sigma)
\langle\mathbf{k}_2|,
\end{equation}
where 
$\phi_\mathbf{k}(\mathbf{r},\sigma)=\langle\mathbf{r},\sigma|\mathbf{k}\rangle$ 
is the wave function of the Bloch state $|\mathbf{k}\rangle$ defined in 
Eq.~\eqref{eq:Bloch-def}, and $\mathbf{k}_1,\mathbf{k}_2$ are each summed 
over a full BZ.~\footnote{Any BZ 
choice is fine, and the two BZs for $\mathbf{k}_1$ and $\mathbf{k}_2$ do not 
have to be the same. It is easy to see that although $|\mathbf{k}\rangle$ is 
only quasi-periodic in $k_y$, $\rho(\mathbf{r},\sigma)$ does not depend on 
the choice of BZ for $\mathbf{k}_1$ or $\mathbf{k}_2$, thanks to the 
quasi-periodicity condition in Eq.~\eqref{eq:Bloch-periodicity}.}
Since $\rho(\mathbf{r},\sigma)$ must have torus periodicity, we can express it as a 
Fourier sum,
\begin{equation}
\rho(\mathbf{r},\sigma)=\frac{1}{L_xL_y}\sum_{\mathbf{q}}\,
e^{i\mathbf{q}\cdot\mathbf{r}}\rho_{\mathbf{q},\sigma}.
\end{equation}
Here, the wave vector $\mathbf{q}$ lives on the reciprocal lattice
\begin{equation}
\mathbf{q}=\left(\frac{2\pi q_x}{L_x},\frac{2\pi q_y}{L_y}\right),\quad
(q_x,q_y)\in\mathbb{Z}^2.
\end{equation}
The projected density operator in momentum space for a single color component 
$\sigma$ is thus given by
\begin{equation}\label{eq:rho-q-integral}
\rho_{\mathbf{q},\sigma}=\!
\sum_{\mathbf{k}_1}^\mathrm{BZ}\!\sum_{\mathbf{k}_2}^\mathrm{BZ}
|\mathbf{k}_1\rangle\langle\mathbf{k}_2|\!
\int\!\mathrm{d}\mathbf{r}\,
e^{-i\mathbf{q}\cdot\mathbf{r}}
\phi^*_{\mathbf{k}_1}\!(\mathbf{r},\sigma)
\phi_{\mathbf{k}_2}\!(\mathbf{r},\sigma),
\end{equation}
where $\int\mathrm{d}\mathbf{r}$ is over the torus $[0,L_x)\times[0,L_y)$.
We define the full projected density operator $\rho_\mathbf{q}$ by
\begin{equation}\label{eq:rho-q-def}
\rho_\mathbf{q}=\sum_\sigma^C \rho_{\mathbf{q},\sigma}.
\end{equation}
This operator is the building block of a color-neutral interacting Hamiltonian.
In Appendix~\ref{sec:density-Bloch}, we finish the integral in 
Eq.~\eqref{eq:rho-q-integral} with the help of the sum over color $\sigma$, 
and prove the main result of this section,
\begin{equation}\label{eq:density-Bloch}
\rho_\mathbf{q}
=e^{-\mathbf{q}^2l_B^2/4}
\sum_{\mathbf{k}}^{\mathrm{BZ}}
e^{-i2\pi q_x(k_y+q_y/2)/N_\phi}\,
|\mathbf{k}\rangle\langle{\mathbf{k}+\mathbf{q}}|.
\end{equation}
It should be noted that when $N_x$ is divisible by $C$, the integral in 
Eq.~\eqref{eq:rho-q-integral} can be finished for each $\sigma$ individually,
without the color sum.
The above formula can be recast [using Eqs.~\eqref{eq:Nf-def} 
and~\eqref{eq:CNf-NxNy}] as
\begin{multline}
\rho_\mathbf{q}=\sum_\mathbf{k}^{\mathrm{BZ}}
|\mathbf{k}\rangle\langle{\mathbf{k}+\mathbf{q}}|\\
\left\{
\exp\left[\frac{\pi}{2}\frac{L_xL_y}{N_xN_y}
\!\left(\!\frac{q_x^2}{L_x^2}+\frac{q_y^2}{L_y^2}\!\right)\!
-i2\pi\frac{q_x(k_y+q_y/2)}{N_xN_y}
\right]\right\}^C.
\end{multline}
Note that the dependence on $C$ enters only through 
the exponent shared by all $\rho_\mathbf{q}$ and all terms in 
$\sum_\mathbf{k}$.

\subsection{Geometric Phase Structure}

The above result suggests that the torus formed by the Bloch states 
$|\mathbf{k}\rangle$ is endowed with a rich geometric structure.
As usual, the Berry connection between the BZ points $\mathbf{k}$ and 
$\mathbf{k}+\mathbf{q}$ is defined as (the phase of) the inner product between 
the periodic part of the Bloch states $|\mathbf{k}\rangle$ and 
$|\mathbf{k}+\mathbf{q}\rangle$. This amounts to the matrix element of 
the operator $e^{-i\mathbf{q}\cdot\mathbf{\hat{r}}}$ between the two states,
where $\mathbf{\hat{r}}$ is the position operator.
Notice that this exponentiated position operator, when projected to the lowest 
Landau level, is nothing but the full density operator $\rho_\mathbf{q}$ in 
Eq.~\eqref{eq:rho-q-def}.
Therefore, we can interpret Eq.~\eqref{eq:density-Bloch} as the parallel 
transport in the momentum space implemented by the projected density 
$\rho_\mathbf{q}$.

Define the primitive vectors on the reciprocal lattice 
$\mathbf{g}_x=(2\pi/L_x,0)$ and $\mathbf{g}_y=(0,2\pi/L_y)$, and the 
shorthand notations $\rho_\alpha=\rho_{\mathbf{q}=\mathbf{g}_\alpha}$ and 
$\mathrm{Phase}[z]=z/|z|$ for $z\in\mathbb{C}$.
At momentum transfer $\mathbf{q}=\mathbf{g}_\alpha$, 
the (unitary) exponentiated Berry connection resolves the band geometry,
\begin{equation}
\mathcal{A}_\alpha(\mathbf{k})
\equiv\mathrm{Phase}[\langle\mathbf{k}|
\rho_\alpha|\mathbf{k}+\mathbf{g}_\alpha\rangle]
=e^{-i2\pi q_x(k_y+q_y/2)/N_\phi},
\end{equation}
while the norm
\begin{equation}\label{eq:quantum-distance}
\Big|\langle\mathbf{k}|\rho_\alpha|\mathbf{k}+\mathbf{g}_\alpha\rangle\Big|
=e^{-\mathbf{q}^2l_B^2/4}
\end{equation}
is the quantum distance between $\mathbf{k}$ and $\mathbf{k}+\mathbf{g}_\alpha$.
Notice that the quantum distance does not depend on $\mathbf{k}$.\footnote{This 
is particular to the Landau level problem; in the tight-binding situation, 
both the quantum distance and the Berry phase depend on $\mathbf{k}$.}
The gauge-invariant Berry phases can be extracted from parallel transport
around closed loops of $|\mathbf{k}\rangle$ states over the BZ torus.

Given that we are interested in the Abelian Berry connection,
each contractible loop
can be decomposed into a product of loops around single plaquettes. 
Such plaquette Wilson loops take a particularly nice form for 
the Bloch states we constructed.
Around the plaquette at $\mathbf{k}$,
\begin{equation}\label{eq:Wsq-def}
\Wsq(\mathbf{k})
\equiv\mathrm{Phase}[\langle\mathbf{k}|
\rho_x\rho_y[\rho_y\rho_x]^{-1}|\mathbf{k}\rangle]
=e^{i2\pi/N_\phi}
\end{equation}
is independent from $\mathbf{k}$.
Further, we can define the Berry curvature over a single plaquette~\cite{Wu13:Bloch}
$f_\mathbf{k}=\frac{1}{2\pi}\Im\log\Wsq(\mathbf{k})$,
where $\Im$ takes the imaginary part in the principal branch 
$\Im\log z\in(-\pi,\pi]$.
We find that the BZ torus for the multicomponent Landau level has constant 
Berry curvature
\begin{equation}
f_\mathbf{k}=\frac{1}{N_\phi},
\end{equation}
and its Chern number is equal to the number of components
\begin{equation}
\sum_\mathbf{k}^{\mathrm{BZ}}f_\mathbf{k}=\frac{N_xN_y}{N_\phi}=C.
\end{equation}

In addition to the contractible loops, there are two independent 
non-contractible Wilson loops around the two fundamental cycles of the torus, 
related to charge polarization.
We define
\begin{equation}
\begin{aligned}
W_x(k_y)&\equiv\mathrm{Phase}[\langle 0,k_y|\rho_x^{N_x}|0,k_y\rangle]
=e^{-i2\pi k_y C/N_y},\\
W_y(k_x)&\equiv\mathrm{Phase}[\langle k_x,0|\rho_y^{N_y}|k_x,0\rangle]
=e^{i2\pi k_x C/N_x}.
\end{aligned}
\end{equation}
The geometric phases over the BZ torus are fully specified by the 
following quantities
\begin{equation}
\{\Wsq(\mathbf{k})|\mathbf{k}\in\mathrm{BZ}\},W_x(0),W_y(0).
\end{equation}
For example, $W_x(1)$ can be obtained from $W_x(0)$ times the product of
$\Wsq(\mathbf{k})$ around each of the $N_x$ plaquettes between $k_y=0$ and 
$k_y=1$ in the first BZ.

We can easily add a twist to the color-entangled boundary conditions in
Eq.~\eqref{eq:color-entangled-boundaries},
\begin{equation}
\begin{aligned}
T(\mathbf{L}_x)P^{N_x}&=e^{-i2\pi\gamma_x},&
T(\mathbf{L}_y)Q^{N_y}&=e^{-i2\pi\gamma_y}.
\end{aligned}
\end{equation}
The twist angles $\boldsymbol{\gamma}=(\gamma_x,\gamma_y)\in\mathbb{R}^2$ 
implement color-independent magnetic flux insertions.
We incorporate this change by keeping $(k_x,k_y)\in\mathbb{Z}^2$, but 
applying
\begin{equation}
\mathbf{k}\rightarrow \mathbf{k}+\boldsymbol{\gamma}
\end{equation}
to every equation so far.

\subsection{Twisted Torus}

The above results can be directly generalized to a twisted torus.
Instead of the rectangular torus spanned by $\mathbf{L}_x=L_x\hat{x}$ and 
$\mathbf{L}_y=L_y\hat{y}$,
we consider a torus with twist angle $\theta$, spanned by
\begin{equation}
\begin{aligned}
\mathbf{L}_x&=L_x\sin\theta\,\hat{x}+L_x\cos\theta\,\hat{y},&
\mathbf{L}_y&=L_y\hat{y}.
\end{aligned}
\end{equation}
The number of fluxes $N_\phi$ is now defined by
\begin{equation}
L_xL_y\sin\theta=2\pi l_B^2 N_\phi.
\end{equation}
The reciprocal lattice primitive vectors $\mathbf{g}_\alpha$ are now defined by
\begin{equation}
\mathbf{g}_\alpha\cdot \mathbf{L}_\beta=2\pi\delta_{\alpha\beta},
\end{equation}
and we have the wave vector $\mathbf{k}=k_x\mathbf{g}_x+k_y\mathbf{g}_y$, 
$(k_x,k_y)\in\mathbb{Z}^2$.
Once we change the wave functions of the hybrid Wannier states in Landau gauge 
$\mathbf{A}=Bx\hat{y}$ to
\begin{multline}
\langle x,y,\sigma|X,k_y\rangle
=\frac{1}{(\sqrt{\pi}L_yl_B)^{1/2}}\!\!
\sum_m^{\mathbb{Z}}
\delta_{\sigma,X+mN_x}^{\text{mod }C}e^{-x^2/(2l_B^2)}\\
\exp\Bigg[
2\pi\!\left(\frac{X N_y+k_y C}{C}+mN_\phi\right)\frac{x+iy}{L_y}\\
-i\pi\frac{L_xe^{-i\theta}}{N_\phi L_y}
\left(\frac{X N_y+k_y C}{C}+mN_\phi\right)^2\Bigg]
,
\end{multline}
all of the earlier results still hold with no essential modifications.
In particular, the proof in Appendix~\ref{sec:density-Bloch} can be adapted 
straight-forwardly (albeit with even more tedious algebra), and 
in Eq.~\eqref{eq:density-Bloch} the density operator requires no formal change 
except for $\mathbf{q}=q_x\mathbf{g}_x+q_y\mathbf{g}_y$.
For the rest of the paper, we return to the rectangular torus. The results 
can be similarly generalized to the twisted torus by simple substitutions.

\section{Pseudopotential Hamiltonian}

With the one-body Bloch and hybrid Wannier bases at hand, we move to the 
many-body interacting problem.
Our ultimate purpose is to build pseudopotential Hamiltonians
for FCI with arbitrary Chern number $C$.
As demonstrated in the last section, the multicomponent LLL 
resembles the Chern band once we impose appropriate boundary conditions 
that join together the $C$ components.
This link enables us to take advantage of the well-developed pseudopotential 
formalism in the LLL.
We construct pseudopotential Hamiltonians (in the same way as those of 
single-component LLL~\cite{Haldane83:Sphere,Simon07:Pseudopotential})
in the LLL from the projected density operator $\rho_\mathbf{q}$,
and obtain its zero modes through numerical diagonalization.
Following the usual practice in the FQH literature,\footnote{For example, the 
Laughlin states at $\nu=1/3$ on a torus can be defined as the exact zero modes 
of the LLL-projected hollow-core interaction.}
we define these zero modes at the FCI model wave functions.

Then, through the mapping between the Bloch states in the LLL and on the 
lattice, we transcribe these LLL wave functions to the lattice.
The resulting trial wave functions can be directly compared with the FCI 
ground states obtained numerically for lattice Hamiltonians.
As demonstrated in our earlier paper~\cite{Wu13:Bloch}, this approach yields 
model Hamiltonians adiabatically connected to the microscopic lattice 
Hamiltonian, and leads to trial wave functions with the correct total momentum 
on lattice and very high overlaps with the actual FCI ground states.
Our trial wave functions also reproduce the anomalous particle entanglement 
spectrum as observed in Ref.~\onlinecite{Sterdyniak13:Higher}.

The question remains, however, how to predict the total lattice momentum for 
the trial wave functions (including quasiholes) \emph{without} numerical 
diagonalization, similar to the methods developed for the 
FQH~\cite{Bernevig08:Jack,Bernevig08:Jack2}.
For $C=1$, this problem was solved by two of us~\cite{Bernevig12:Counting} by 
combining the generalized Pauli principle~\cite{Bernevig08:Jack,Bernevig08:Jack2}
for single-component FQH states (including quasiholes) with lattice folding.
For $C>1$, we now have the LLL-to-lattice mapping. What we still lack is a 
multicomponent version of the generalized Pauli principle.
Refs.~\onlinecite{Estienne12:Singlet,Ardonne11:Squeezing} studied this 
problem for the usual boundary conditions.
Due to our modifications to the boundary conditions, their 
results do not directly apply here.

Fortunately, we can also extract the generalized Pauli principle from 
the Hamiltonian in the thin-torus 
limit~\cite{Tao83:TT,Bergholtz05:TT,Seidel06:TT}.
In this limit, the hybrid Wannier orbitals in the LLL become isolated from 
each other.
Specifically, we find from Eq.~\eqref{eq:Wannier-wf} 
that the ratio between the width of the hybrid Wannier orbital and the spacing 
between them scales as
\begin{equation}
\frac{\mathrm{width}}{\mathrm{spacing}}\sim \frac{l_B}{2\pi l_B^2/L_y}
\sim \sqrt{N_\phi \frac{L_y}{L_x}}.
\end{equation}
Therefore, when the aspect ratio $L_x/L_y$ satisfies
\begin{equation}\label{eq:thin-torus}
\frac{L_x}{L_y}\gg N_\phi,
\end{equation}
the hybrid Wannier orbitals are so separated that the projected density 
operator becomes approximately diagonal in the hybrid Wannier basis.
As a result, the pseudopotential Hamiltonian built from projected density 
operators also becomes approximately diagonal in the hybrid Wannier basis.
(This is not true for certain non-unitary states~\cite{Papic13:TT}.)
By analyzing the classical electrostatics of the leading terms in the 
Hamiltonian, we can obtain the quantum numbers of the Hamiltonian
zero modes.
(For FQH with the usual boundary conditions, this was done in 
Refs.~\onlinecite{Bergholtz05:TT,Bergholtz06:TT,Bergholtz08:TTCounting}.)
After the Bloch mapping between FCI and FQH, this will give us a counting rule 
for the degeneracy of the FCI quasiholes in each lattice momentum sector.

In the rest of this Section, we expand the new pseudopotential Hamiltonian 
proposed earlier~\cite{Wu13:Bloch} in the Wannier basis, and perform the 
necessary resummation to make it amenable to proper truncation in the 
thin-torus limit. The actual truncation and the analysis of the zero modes of 
the truncated Hamiltonian is left for the next Section.

\subsection{Projected Density in the Hybrid Wannier Basis}

We obtain the projected density operator in the hybrid Wannier basis by 
plugging the Fourier transform Eq.~\eqref{eq:Bloch-def} into 
Eq.~\eqref{eq:density-Bloch},
\begin{multline}\label{eq:density-Wannier}
\rho_\mathbf{q}
=e^{-\mathbf{q}^2l_B^2/4}
\sum_{X}^{N_x}
\sum_{k_y}^{N_y}
e^{-i2\pi q_x[(XN_y+k_yC)/C+q_y/2]/N_\phi}\\
|X,k_y\rangle\langle X,k_y+q_y|.
\end{multline}
Notice that the phase factor depends on the summation variables $X,k_y$ only 
through the linear combination $XN_y+k_yC$, which is proportional to the 
center position of the hybrid Wannier orbital
$|X,k_y\rangle$ [Eq.~\eqref{eq:Wannier-wf}],
\begin{equation}
\langle X,k_y|\hat{x}|X,k_y\rangle
=L_x\frac{XN_y+k_yC}{N_xN_y}\text{ mod }L_x,
\end{equation}
where $\hat{x}$ is the position operator in the $x$ direction.
This motivates us to index these orbitals by their center position.
In the following, we introduce an alternative labeling $|j,s\rangle$ for 
the Wannier states. The $j$ index gives the center position of the Wannier
state while the $s$ index plays a role similar (but not identical)
to the color index $\sigma$.
As we will see in the next Section, the projected interaction decays
exponentially when the difference in the $j$ indices between two particles 
increases.

As seen from Eq.~\eqref{eq:Wannier-wf}, the hybrid Wannier state
$|X,k_y\rangle$ depends on $(X,k_y)$ only through 
\begin{equation}
XN_y+k_yC
\end{equation}
and
\begin{equation}
X\text{ mod }C,
\end{equation}
in the exponential and the Kronecker-$\delta$ in Eq.~\eqref{eq:Wannier-wf}, 
respectively.
For integers $X,k_y$, the linear combination $XN_y+k_yC$ 
must be an integer multiple of the greatest common divisor (GCD)
\begin{equation}\label{eq:Ct-def}
\widetilde{C}\equiv \mathrm{GCD}(C,N_y).
\end{equation}
Therefore, we introduce two integer labels
\begin{equation}\label{eq:j-s-def}
\begin{aligned}
j&=(XN_y+k_yC)/\widetilde{C},\\
s&=X\text{ mod }C.
\end{aligned}
\end{equation}
For future convenience, we also define integers
\begin{equation}\label{eq:M-d-def}
\begin{aligned}
M&=N_xN_y/\widetilde{C},\\
d&=C/\widetilde{C},
\end{aligned}
\end{equation}

\begin{figure}[tb]
\centering
\includegraphics[]{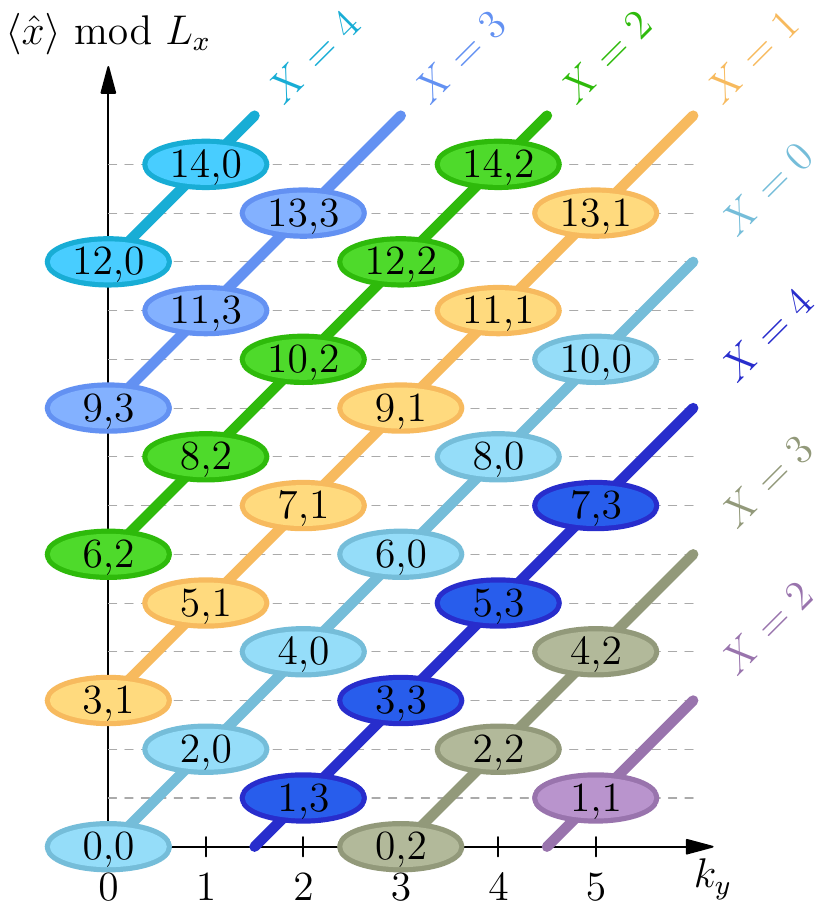}%
\caption{\label{fig:Wannier-relabeling}
Relabeling of the Wannier states $|X,k_y\rangle\leftrightarrow|j,s\rangle$
for $(N_x,N_y)=(5,6)$ and $C=4$.
We focus on the principal region with $X\in[{0~..~N_x})$ and 
$k_y\in[{0~..~N_y})$.
Each solid ellipse represents a Wannier center.
The horizontal axis gives the $k_y$ index while the vertical
axis gives the position of the Wannier center in the $x$ direction (mod $L_x$).
The ellipses are colored according to the $X$ index, and labeled
by the $(j,s)$ indices.
We have employed Eq.~\eqref{eq:Wannier-periodicity-relabeled} to shift $j$ to 
$[0~..~M)$ and $s$ to $[0~..~C)$.
Upon a color-independent flux insertion,
each Wannier center flows along the solid lines of its color.
}
\end{figure}

We emphasize that $j$ and $s$ are not independent.
This can be seen by examining the solutions to the first equation in 
Eq.~\eqref{eq:j-s-def}.
For a given $j$, if $(X,k_y)$ is a solution, then all the solutions can 
be parametrized as $(X+nC/\widetilde{C}, k_y - nN_y/\widetilde{C})$, 
$n\in\mathbb{Z}$. Therefore $s=X$ mod $C$ can take $\widetilde{C}$ 
different values in $[0~..~C)$ with uniform spacing $d=C/\widetilde{C}$
[Eq.~\eqref{eq:M-d-def}], corresponding to $n=1,\ldots,\widetilde{C}$ in 
$X+nC/\widetilde{C}$.
For a given $j$,
we denote this set of $\widetilde{C}$ allowed values of $s$ by
\begin{equation}\label{eq:Sj}
S_j\subset [0~..~C).
\end{equation}
A useful property is
\begin{equation}\label{eq:Sj-d}
S_j=S_{j+d},
\end{equation}
which follows from the fact that $j\rightarrow j+d$ can be achieved 
by $k_y\rightarrow k_y+1$ without touching $X$.
Plugging Eq.~\eqref{eq:j-s-def} into Eq.~\eqref{eq:Wannier-wf}, we find that 
indeed we can relabel the hybrid Wannier states
\begin{equation}\label{eq:Wannier-relabeling}
|X,k_y\rangle \leftrightarrow |j,s\rangle,
\end{equation}
modulo the identification
\begin{equation}
|j,s\rangle=|j,s+C\rangle.
\end{equation}
An example is given in Fig.~\ref{fig:Wannier-relabeling}.
It is not hard to see that this mapping is bijective,
although we cannot easily write down an explicit formula for the solution 
$(X,k_y)$ to Eq.~\eqref{eq:j-s-def} at a given $(j,s)$.
We denote the $k_y$ solution formally as
\begin{equation}
k_y(j,s).
\end{equation}
Then, the representation of the color-entangled magnetic translations 
$\widetilde{T}_\alpha$ in the $|j,s\rangle$ basis can be constructed 
indirectly from Eq.~\eqref{eq:Wannier-translations},
\begin{equation}\label{eq:Wannier-translations-relabeled}
\begin{aligned}
\widetilde{T}_x|j,s\rangle&=|j+N_y/\widetilde{C},s+1\rangle,\\
\widetilde{T}_y|j,s\rangle&=e^{-i2\pi k_y(j,s)/N_y}|j,s\rangle.
\end{aligned}
\end{equation}
The wave functions for $|j,s\rangle$ can be obtained from 
Eq.~\eqref{eq:Wannier-wf},
\begin{multline}\label{eq:Wannier-wf-relabeled}
\langle x,y,\sigma|j,s\rangle
=\frac{1}{(\sqrt{\pi}L_yl_B)^{1/2}}
\sum_m^{\mathbb{Z}}
\delta_{\sigma,s+mN_x}^{\text{mod }C}\\
\exp\Big\{\textstyle
i2\pi\!\left(\frac{j}{d}+mN_\phi\right)\frac{y}{L_y}
-\frac{1}{2}\!\left[
\frac{x}{l_B}\!-\!\frac{2\pi l_B}{L_y}\left(\frac{j}{d}+mN_\phi\right)
\right]^2\!
\Big\}.
\end{multline}
In parallel to Eq.~\eqref{eq:Wannier-periodicity},
$|j,s\rangle$ is periodic in $s$ but quasi-periodic in $j$,
\begin{equation}\label{eq:Wannier-periodicity-relabeled}
\begin{aligned}
|j+M,s+N_x\rangle&=|j,s\rangle,\\
|j,s+C\rangle&=|j,s\rangle.
\end{aligned}
\end{equation}
As we will see soon, this twist in $s$ when shifting $j$ is the main issue
that sets the current problem apart from the usual multicomponent FQH.~\cite{Estienne12:Singlet}

We now want to expand the projected density operator in the relabeled hybrid 
Wannier basis.
On the one hand, notice that due to the quasi-periodicity of $|X,k_y\rangle$ 
[Eq.~\eqref{eq:Wannier-periodicity}], the double sum of $(X,k_y)$ over 
$[0~..~N_x)\times [0~..~N_y)$ in Eq.~\eqref{eq:density-Wannier} can be shifted 
to any set of $N_xN_y$ points in the $\mathbb{Z}^2$ plane, as long as the 
corresponding hybrid states are independent from each other.
On the other hand, notice that
\begin{equation}\label{eq:j-s-range}
\Big\{\,|j,s\rangle \,\Big|\,j\in[j_0~..~j_0+M), s\in S_j\Big\}
\end{equation}
label a set of $N_xN_y$ hybrid Wannier states that are independent from each 
other for any given $j_0\in\mathbb{Z}$.
Therefore, we can rewrite the double sum in Eq.~\eqref{eq:density-Wannier} as 
a sum over the above set.
Since increasing $k_y$ by $q_y$ while keeping $X$ constant sends $(j,s)$ to 
$(j+q_yd, s)$, we have 
\begin{equation}\label{eq:density-Wannier-relabeled}
\rho_\mathbf{q}
=e^{-\mathbf{q}^2l_B^2/4}
\sum_j^{M}\!{}'
e^{-i2\pi q_x(j+\frac{q_yd}{2})/M}
\sum_s^{S_j}
|j,s\rangle\langle j+q_yd,s|,
\end{equation}
where the primed sum is over $j\in[j_0~..~j_0+M)$ for an arbitrary 
$j_0\in\mathbb{Z}$, with $M=N_xN_y/\widetilde{C}$ [Eq.~\eqref{eq:M-d-def}].
The appearance of $\langle j+q_yd,s|$ requires special attention:
when we shift $j+q_yd$ back to $[j_0~..~j_0+M)$ using 
Eq.~\eqref{eq:Wannier-periodicity-relabeled}, the $s$ index must be changed 
accordingly, by $N_x$ (mod $C$).
This boundary effect dictates that $\rho_\mathbf{q}$ is \emph{not} diagonal in 
$s$ unless $N_x$ is divisible by $C$,
which discourages a seemingly plausible interpretation of $s$ as an effective 
spin index in general.

\subsection{Interacting Hamiltonian}
We consider only interactions between a pair of
color-neutral densities $\rho_\mathbf{q}$.
The relevance of such interactions to the Chern insulators was justified 
numerically in our previous paper~\cite{Wu13:Bloch}.
Such interactions can be specified in terms of the Haldane pseudopotentials.
Higher-body pseudopotentials\cite{Simon07:Pseudopotential}
can be implemented in the same spirit.
We consider only the first two pseudopotentials $(V_0,V_1)$ 
being non-negative, with all $V_{m>1}=0$.
The interaction strength at momentum transfer $\mathbf{q}$ then reads
\begin{equation}\label{eq:V0-V1}
V_\mathbf{q}=4\pi l_B^2 [V_0+V_1\cdot (1-\mathbf{q}^2l_B^2)],
\end{equation}
and the Hamiltonian is given by
\begin{equation}\label{eq:H}
H=\frac{1}{2L_xL_y}
\sum_\mathbf{q} V_\mathbf{q} \rho_\mathbf{q}\rho_{-\mathbf{q}}.
\end{equation}
Here $\mathbf{q}$ is summed over the infinite reciprocal lattice.

As shown in our previous paper~\cite{Wu13:Bloch}, the color-entangled 
generalizations of the bosonic/fermionic Halperin singlet states and the 
corresponding quasihole states are defined as the exact zero modes of the 
above Hamiltonian (using $V_1=0$ for the bosonic case).
These states are distinct from the usual Halperin states due to the 
color-entangled boundary conditions inherent in $\rho_\mathbf{q}$.
Through numerical diagonalization, we can obtain these zero modes, and then 
transcribe them to the lattice system of an arbitrary Chern insulator using 
the one-body mapping between the LLL Bloch states and the lattice Bloch states.
We now attempt to achieve an analytic understanding of this Hamiltonian, by 
exploiting its assumed adiabatic connectivity~\cite{Wu13:Bloch} to the 
thin-torus limit.

We first plug Eq.~\eqref{eq:density-Wannier-relabeled} into Eq.~\eqref{eq:H} 
and write $H$ in the relabeled hybrid Wannier basis,
\begin{multline}\label{eq:H-Wannier}
\!\!H\!=\frac{1}{2L_xL_y}\!\!\sum_\mathbf{q}e^{-\mathbf{q}^2l_B^2/2}
V_\mathbf{q}
\sum_{j_1}^M\!{}'\sum_{j_2}^M\!{}'
e^{-i2\pi q_x(j_1-j_2+q_yd)/M}\\
\sum_{s_1}^{S_{j_1}}\!
\sum_{s_2}^{S_{j_2}}
\psi^\dagger_{j_1,s_1}
\psi^\dagger_{j_2,s_2}
\psi_{j_2-q_yd,s_2}
\psi_{j_1+q_yd,s_1},
\end{multline}
where $M$ and $d$ are defined in Eq.~\eqref{eq:M-d-def}, and
for $\mathbf{q}=(2\pi q_x/L_x,2\pi q_y/L_y)$, we have
\begin{equation}
\mathbf{q}^2l_B^2=\frac{2\pi}{N_\phi}
\left(\frac{L_y}{L_x}q_x^2+\frac{L_x}{L_y}q_y^2\right).
\end{equation}
We want to massage the above expansion of $H$ to a form amenable to justified 
truncation in the thin-torus limit.
The main obstacle is obviously the oscillatory factor 
$e^{-i2\pi q_x(j_1-j_2+q_yd)/M}$ in the coefficient.
This can be removed in exchange for a Gaussian factor by performing a Poisson 
resummation over $q_x$,
which does not appear in the index of the creation/annihilation operators.
After some straightforward but tedious algebra in Appendix~\ref{sec:sum-qx}, 
we find
\begin{multline}\label{eq:H-resummed}
\!\!\! H=\sqrt{\frac{L_x}{N_\phi L_y}}
\sum_{q_y}^{\mathbb{Z}}
e^{-\beta(q_yd)^2}
\sum_{j}^M\!{}'\sum_{\Delta}
\sum_n^{\mathbb{Z}}
e^{-\beta(\Delta-q_yd+nM)^2}\\
\left\{
V_0 + 2\beta V_1\left[(\Delta-q_yd+nM)^2-(q_yd)^2\right]
\right\}\\
\sum_{s}^{S_{j}}
\sum_{s'}^{S_{j+\Delta}}
\psi^\dagger_{j,s}
\psi^\dagger_{j+\Delta,s'}
\psi_{j+\Delta-q_yd,s'}
\psi_{j+q_yd,s},
\end{multline}
where $\Delta$ is summed over an interval of length $M$ centered around $q_yd$,
\begin{equation}\label{eq:delta-range}
\Delta\in\big[\,q_yd-\lfloor M/2\rfloor~..~q_yd-\lfloor M/2\rfloor+M\,\big),
\end{equation}
and we have defined the shorthand
\begin{equation}
\beta=\frac{1}{d^2}\frac{\pi}{N_\phi}\frac{L_x}{L_y}.
\end{equation}

\section{Thin-Torus Analysis}

\begin{figure}[tb]
\centering
\includegraphics[]{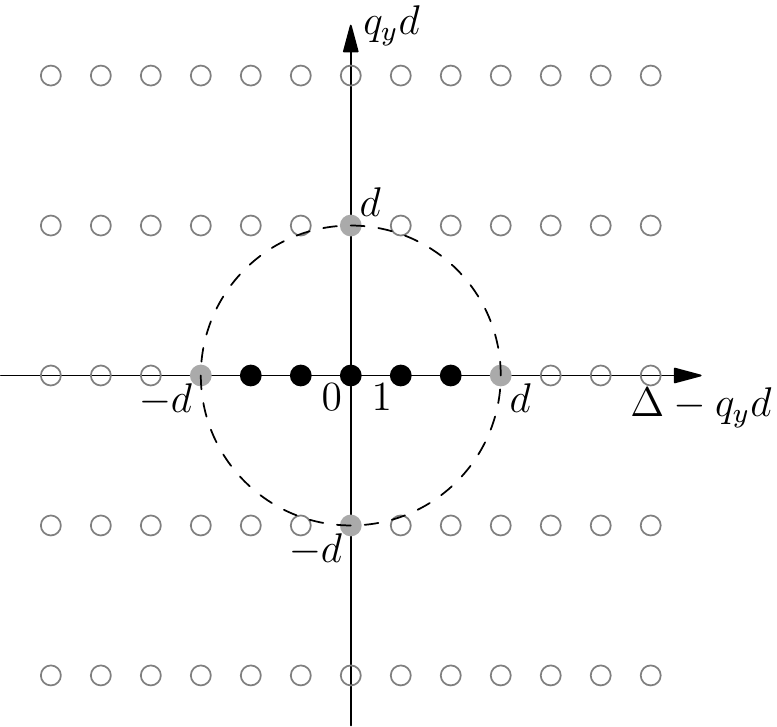}%
\caption{\label{fig:truncation-boson}
Terms in the expansion of the pseudopotential Hamiltonian.
Here we illustrate the example of $d=3$.
Each dot represents a term $(q_y, \Delta)$ in Eq.~\eqref{eq:H-resummed-boson}.
The weight of each term decays exponentially in its distance from the origin.
The dashed circle marks the empirical threshold for truncation 
$(q_yd)^2+(\Delta-q_yd)^2=d^2$.
The solid black dots inside are the density-density terms in 
Eq.~\eqref{eq:truncation-boson-density-terms}, while the four solid gray dots 
contain the pair hopping and the density-density terms in 
Eqs.~\eqref{eq:truncation-boson-mixed-terms-d1} 
and~\eqref{eq:truncation-boson-mixed-terms}.
}
\end{figure}

In Eq.~\eqref{eq:H-resummed}, the Hamiltonian has been organized into groups 
of density-density or pair hopping terms.
The strengths of the terms decay exponentially in the limit
\begin{equation}
\beta\gg 1.
\end{equation}
This is exactly the thin-torus limit in Eq.~\eqref{eq:thin-torus}.
In the following, we perform a proper truncation of the Hamiltonian in this 
limit and analyze the degeneracy and quantum numbers of its zero modes. 

The thin-torus analysis is a well-known, powerful technique to tackle the 
strongly-correlated physics in single-component FQH 
effect.~\cite{Bergholtz05:TT,Bergholtz06:TT,Bergholtz08:TTCounting,Bernevig12:TT}
In the thin-torus limit, the pair hopping terms die off quickly,
and the Hamiltonian becomes classical, dominated
by density-density terms and thus solvable.
(This is not true for certain non-unitary states~\cite{Papic13:TT}.)
One can obtain the correct degeneracy of the ground states and extract their 
total momenta simply by minimizing the classical electrostatic energy and 
completely ignoring the pair hoppings.
By assumed adiabatic connectivity,~\cite{Wu13:Bloch}
the results must also apply to the isotropic limit.
The thin-torus analysis thus provides an intuitive picture for the `root 
partitions' and the underlying generalized Pauli 
principle of Refs.~\onlinecite{Bernevig08:Jack,Bernevig08:Jack2}.
Our multicomponent pseudopotential Hamiltonian with color-entangled 
boundaries~\eqref{eq:H-resummed} turns out to be considerably more 
complicated, due to the essential role played by the pair hopping terms.
As we will see soon, the largest pair hopping terms have strengths 
comparable to the subleading density-density terms.
Keeping only the leading density-density terms results in too many zero modes 
compared with the numerical
studies~\cite{Sterdyniak13:Higher,Liu12:Higher,Wu13:Bloch}.
The correct ground-state degeneracy is recovered only after we put back the 
largest pair hoppings, which turn out to be of similar strength as some of 
the density-density terms.
This indicates that the thin-torus limit of our multicomponent pseudopotential 
Hamiltonian cannot be described by classical electrostatics alone.
The useful result of this Section is a set of rules 
[Sec.~\ref{sec:counting-rule}] that 
correctly predict the degeneracy and total lattice momenta of FCI ground 
states (including quasiholes). This is illustrated by explicit examples in 
Secs.~\ref{sec:counting-example} and~\ref{sec:counting-example-qh}.

\subsection{Truncation of Bosonic Hamiltonian}

Numerical studies in Refs.~\onlinecite{Sterdyniak13:Higher,Liu12:Higher} found 
gapped FCI phases of bosons at filling $\nu=1/(C+1)$ with $(C+1)$-fold 
degenerate ground states, stabilized by on-site interactions projected to a 
topological flat band with Chern number $C$.
In the following we specialize to the simplest case of bosons
and try to understand the ground states of the 
pseudopotential Hamiltonian at filling $\nu=1/(C+1)$ and with quasiholes.
Setting $V_0=\sqrt{N_\phi L_y/L_x}>0$ and $V_1=0$, the Hamiltonian in 
Eq.~\eqref{eq:H-resummed} becomes
\begin{multline}\label{eq:H-resummed-boson}
H=\sum_{q_y}^{\mathbb{Z}}
\sum_{j}^M\!{}'\sum_{\Delta}
\sum_n^{\mathbb{Z}}
e^{-\beta(q_yd)^2-\beta(\Delta-q_yd+nM)^2}\\
\sum_{s}^{S_{j}}
\sum_{s'}^{S_{j+\Delta}}
\psi^\dagger_{j,s}
\psi^\dagger_{j+\Delta,s'}
\psi_{j+\Delta-q_yd,s'}
\psi_{j+q_yd,s}.
\end{multline}
where the primed sum of $j$ is over
\begin{equation}\label{eq:j-range}
j\in[j_0~..~j_0+M)
\end{equation}
for an arbitrary $j_0\in\mathbb{Z}$ [Eq.~\eqref{eq:j-s-range}], while
$\Delta$ is summed over the interval of length $M$
given in Eq.~\eqref{eq:delta-range}.

In the $\beta\gg1$ limit, we can safely truncate the sum over $n$ to a single 
term at $n=0$, if we assume that $M/d=N_\phi\gg 1$.
Further, only the terms with $q_y\sim0$ and $\Delta-q_yd\sim0$ have a 
significant contribution, since the coefficients decay exponentially with 
respect to the (squared) Euclidean distance from $q_yd=\Delta-q_yd=0$,
\begin{equation}
R^2(q_y,\Delta) \equiv (q_yd)^2+(\Delta-q_yd)^2,
\end{equation}
as illustrated in Fig.~\ref{fig:truncation-boson}.
The 4-boson $\psi^\dagger\psi^\dagger\psi\psi$ operator can be either 
density-density interaction or pair hopping.
We find that the terms with $q_y=0$ are all density-density interactions,
while the strongest pair hopping terms may appear at $|q_y|=1$, $\Delta=q_yd$,
with Euclidean distance $R^2=d^2$.

In light of the previous 
studies~\cite{Bergholtz05:TT,Bergholtz06:TT,Bergholtz08:TTCounting,Bernevig12:TT},
we first examine the effect of the terms with
$R^2(q_y,\Delta)<d^2$.
They can be collected into
\begin{equation}\label{eq:truncation-boson-density-terms}
H_{<d^2}=\sum_j^M
\sum_\Delta^{(-d~..~d)}e^{-\beta \Delta^2}n_j n_{j+\Delta},
\end{equation}
where the number operator $n_j$ is defined by
\begin{equation}\label{eq:number-j}
n_j=\sum_s^{S_j}\psi^\dagger_{j,s}\psi_{j,s}.
\end{equation}

Recall from Eq.~\eqref{eq:Sj} that $S_j$ is the set of all allowed values of 
$s$ for $\psi_{j,s}$ at a given $j$, and this set contains $\widetilde{C}$ 
different values. Also, recall from Eq.~\eqref{eq:M-d-def} that 
$d\,\widetilde{C}=C$.
By solving the simple electrostatics, we find that
the zero modes of $H_{<d^2}$ with highest density appear at filling 
$\nu=1/C$. This leads to much more than $(C+1)$ zero modes at
filling $1/(C+1)$, inconsistent with the findings
from numerical diagonalization 
of actual FCI Hamiltonians.~\cite{Sterdyniak13:Higher,Liu12:Higher}.
This is a clear signal that we should include more terms in the truncated 
Hamiltonian.

In the following we analyze the effect of the next strongest terms in
Eq.~\eqref{eq:H-resummed-boson},
with Euclidean distance $R^2(q_y,\Delta)=d^2$.
They are located at $({|\Delta-q_yd|},|q_yd|)=(0,d)$ and $(d,0)$,
represented by the four solid gray dots in Fig.~\ref{fig:truncation-boson}.
In the next section we provide detailed analysis of the simplest case with 
$d=1$.
The results for general $d$ will be presented afterwards.

\subsection{Effect of Pair Hopping Terms: $d=1$}
\label{sec:zero-modes-d1}

\begin{figure}[tb]
\centering
\includegraphics[]{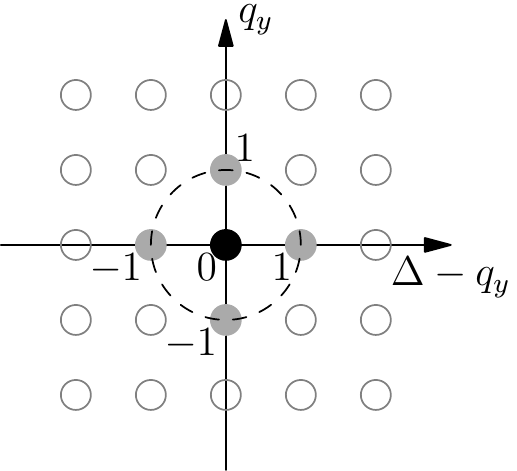}%
\caption{\label{fig:truncation-boson-d1}
Terms in the expansion of the pseudopotential Hamiltonian at $d=1$.
The presentation follows the same format as Fig.~\ref{fig:truncation-boson}.
}
\end{figure}

In this subsection we specialize to the simplest case $d=1$, illustrated in 
Fig.~\eqref{fig:truncation-boson-d1}.
In this case $N_y$ is divisible by $C$ [Eqs.~\eqref{eq:Ct-def} 
and~\eqref{eq:M-d-def}].
The pseudopotential Hamiltonian in 
Eq.~\eqref{eq:H-resummed-boson} (after truncating the sum over $n$) becomes
\begin{multline}\label{eq:H-resummed-boson-d1}
H=\sum_{q_y}^{\mathbb{Z}}
\sum_{j}^M\!{}'\sum_{\Delta}
e^{-\beta q_y^2-\beta(\Delta-q_y)^2}\\
\sum_{s}^{S_{j}}
\sum_{s'}^{S_{j+\Delta}}
\psi^\dagger_{j,s}
\psi^\dagger_{j+\Delta,s'}
\psi_{j+\Delta-q_y,s'}
\psi_{j+q_y,s}.
\end{multline}
We now extract the terms at $q_y^2+(\Delta-q_y)^2=d^2=1$, namely, 
at $(q_y,\Delta)=(1,1),(0,1),(-1,-1),(0,-1)$.
To collect together the terms nicely, recall from Eq.~\eqref{eq:Sj-d} that 
$S_j=S_{j+1}$ at $d=1$, and note that we can take advantage of the freedom in 
Eq.~\eqref{eq:j-range} to shift the range of the primed sum over $j$.
We then find
\begin{equation}\label{eq:truncation-boson-mixed-terms-d1}
\sum_j^M e^{-\beta}
\sum_{s}^{S_j}
\sum_{s'}^{S_j}
\Big[
(1,1)+(0,1)+(-1,-1)+(0,-1)
\Big].
\end{equation}
The four terms in the above brackets are labeled by $(q_y,\Delta)$, and explicitly 
they are given by
\begin{equation}
\begin{aligned}
&\psi^\dagger_{j,s}
\psi^\dagger_{j+1,s'}
\psi_{j,s'}
\psi_{j+1,s}\!
+\psi^\dagger_{j,s}
\psi^\dagger_{j+1,s'}
\psi_{j+1,s'}
\psi_{j,s}\\
+&\psi^\dagger_{j+1,s}
\psi^\dagger_{j,s'}
\psi_{j+1,s'}
\psi_{j,s}\!
+\psi^\dagger_{j+1,s}
\psi^\dagger_{j,s'}
\psi_{j,s'}
\psi_{j+1,s}.
\end{aligned}
\end{equation}
Further, notice that we can combine the above four terms into a single product,
\begin{equation}
b^\dagger_{j,s,s'}
b_{j,s,s'},
\end{equation}
where the pair annihilation operator is given by
\begin{equation}\label{eq:pair-annihilation-d1}
b_{j,s,s'}=
\psi_{j,s'}
\psi_{j+1,s}
+\psi_{j+1,s'}
\psi_{j,s}.
\end{equation}
This combination is the key to the enumeration of zero modes as we detail 
below.
Together with the density-density terms in 
Eq.~\eqref{eq:truncation-boson-density-terms}, the bosonic pseudopotential 
Hamiltonian takes the truncated form
\begin{multline}\label{eq:H-truncated-boson-d1}
H=\sum_j^M\Bigg[
n_j n_j
+e^{-\beta}
\bigg(
2n_j n_{j+1}+
\sum_{s}^{S_j}
\sum_{s'}^{S_j}
b^\dagger_{j,s,s'}
b_{j,s,s'}
\bigg)
\Bigg]\\
+\sum_j^M\,\mathcal{O}\Big(e^{-2\beta}\Big).
\end{multline}
The residual terms are exponentially small for $\beta\gg 1$.

When $\widetilde{C}=1$, the $s$ index can take only a single value $s=0$,
reducing $b_{j,s,s'}$ to $2\psi_{j,0}\psi_{j+1,0}$.
This includes the case of Chern number $C=1$.
The truncated Hamiltonian becomes very simple:
\begin{equation}
H=\sum_j^M
\left(
n_j n_j
+4e^{-\beta}
n_j n_{j+1}\right)
+\sum_j^M\,\mathcal{O}\Big(e^{-2\beta}\Big).
\end{equation}
Its zeros modes have no more than one boson in two consecutive orbitals.
We thus recover the familiar result~\cite{Bergholtz05:TT,Bernevig08:Jack}
for the bosonic Laughlin state at half filling.

We now come back to the case with generic $C$.
We look for the constraints on the zero modes of the above truncated 
Hamiltonian in Eq.~\eqref{eq:H-truncated-boson-d1}.
Due to the two-body nature of the interaction, we only need to consider a pair 
of bosons at a time, with $j$ indices being $j_1,j_2$.
In Eq.~\eqref{eq:H-truncated-boson-d1}, each term in the summation is 
positive-semidefinite by itself. This means that
to find the zero modes of Eq.~\eqref{eq:H-truncated-boson-d1},
we only need to identify 
the null space of each term individually, and then take their intersection.
From the density-density terms, we find that in a zero mode we must have
\begin{equation}\label{eq:dj-bound-d1}
|j_1-j_2|\geq 1.
\end{equation}
This amounts to a minimal distance between adjacent bosons along the $j$ axis, 
with no discrimination of the $s$ indices.
The pair hopping terms $\sum b^\dagger b$ in 
Eq.~\eqref{eq:H-truncated-boson-d1} kick in only when the equality sign 
is taken in Eq.~\eqref{eq:dj-bound-d1}, as is evident from 
Eq.~\eqref{eq:pair-annihilation-d1}.
Specifically, $\sum b^\dagger b$ enforces in a zero mode
the antisymmetrization of the $s$ 
indices between bosons with $|j_1-j_2|=1$,
\begin{equation}\label{eq:s-antisymm}
\left(
\psi^\dagger_{j_1,s_1}\psi^\dagger_{j_2,s_2}
-\psi^\dagger_{j_1,s_2}\psi^\dagger_{j_2,s_1}
\right)\,
\big|\emptyset\big\rangle.
\end{equation}
We emphasize that the $\psi^\dagger$'s are bosonic operators.
It is easy to verify that the above antisymmetrized form is indeed annihilated 
by $\sum b^\dagger b$, whereas the symmetrized form 
acquires a positive energy $2e^{-\beta}$.
To find the zero modes for a system of $N$ bosons,
we need to perform the above procedure on each pair of bosons.
This is explained in more details in Sec.~\ref{sec:counting-rule},
and illustrated by an example in Sec.~\ref{sec:counting-example-qh}.

One last subtlety comes from the quasi-periodicity of the $j$ index 
[Eq.~\eqref{eq:Wannier-periodicity-relabeled}].
The orbitals at $j+M$ are identified with those at $j$, but there is a 
possible mismatch between the $s$ indices,
\begin{equation}\label{eq:Wannier-quasiperiodicity-relabeled}
|j+M,s\rangle=|j,s-N_x\rangle.
\end{equation}
For the density terms, this does not make much trouble since $n_j=n_{j+M}$ 
after the summation of the $s$ index over $S_j$ [Eq.~\eqref{eq:number-j}];
we just need to enforce the minimal distance condition 
[Eq.~\eqref{eq:dj-bound-d1}] across the quasi-periodic boundary $j=0$ mod $M$.
For the pair hopping terms, however, we have to be more careful about the $s$ 
index mismatch.
We have to first shift their $j$ indices (by integer multiples of $M$) such 
that $|j_1-j_2|=1$ \emph{before} we can apply the antisymmetrization in 
Eq.~\eqref{eq:s-antisymm}.
More explicitly, if $|j_1-j_2+M|=1$ for example, then the correct 
antisymmetrization can be either of the following two equivalents,
\begin{equation}\label{eq:s-antisymm-boundary}
\begin{aligned}
\big(
\psi^\dagger_{j_1+M,s_1}\psi^\dagger_{j_2,s_2}
&-\psi^\dagger_{j_1+M,s_2}\psi^\dagger_{j_2,s_1}
\big)\,
\big|\emptyset\big\rangle\\
=\big(
\psi^\dagger_{j_1,s_1-N_x}\psi^\dagger_{j_2,s_2}
&-\psi^\dagger_{j_1,s_2-N_x}\psi^\dagger_{j_2,s_1}
\big)\,
\big|\emptyset\big\rangle,
\end{aligned}
\end{equation}
but \emph{not} Eq.~\eqref{eq:s-antisymm} anymore.
This is the only reason why we were not able to consistently 
implement~\cite{Wu13:Bloch} the exclusion principle for conventional 
multicomponent FQH model states~\cite{Estienne12:Singlet,Ardonne11:Squeezing}
for the color-entangled system.

\subsection{Effect of Pair Hopping Terms: General $d$}
\label{sec:zero-modes-general-d}

The analysis for general $d$ is not much different from $d=1$.
Here we just state the essential results.
The pair hopping and density-density terms at $(q_yd)^2+({\Delta-q_yd})^2=d^2$ 
can be merged together,
\begin{equation}\label{eq:truncation-boson-mixed-terms}
\sum_j^M e^{-\beta d^2}
\sum_{s}^{S_j}
\sum_{s'}^{S_j}
b^\dagger_{j,s,s'}
b_{j,s,s'},
\end{equation}
where the two-body annihilation operator is given by
\begin{equation}\label{eq:pair-annihilation}
b_{j,s,s'}=
\psi_{j,s'}
\psi_{j+d,s}
+\psi_{j+d,s'}
\psi_{j,s}.
\end{equation}
Combined with the density-density terms in 
Eq.~\eqref{eq:truncation-boson-density-terms},
the leading terms in the bosonic pseudopotential Hamiltonian in the limit of 
$\beta\gg 1$ are
\begin{multline}\label{eq:H-truncated-boson}
H=\sum_j^M\Bigg[
\sum_\Delta^{(-d~..~d)}e^{-\beta \Delta^2}n_j n_{j+\Delta}\\
+e^{-\beta d^2}
\sum_{s}^{S_j}
\sum_{s'}^{S_j}
b^\dagger_{j,s,s'}
b_{j,s,s'}
\Bigg]\\
+\sum_j^M\,\mathcal{O}\Big(e^{-\beta (d^2+1)}\Big).
\end{multline}
The zero modes of the truncated Hamiltonian satisfy the following pairwise 
constraints.
First, for a pair of bosons with $j$ indices being $j_1$ and $j_2$, we must have
\begin{equation}\label{eq:dj-bound}
|j_1-j_2|\geq d.
\end{equation}
Here the difference in $j$ is understood with the quasi-periodic 
identification $j\sim j+M$.
When the equality in Eq.~\eqref{eq:dj-bound} holds, the two bosons are further 
subject to an antisymmetrization in the $s$ indices.
For the simplest case $|j_1-j_2|=d$, we need Eq.~\eqref{eq:s-antisymm},
whereas for $|j_1-j_2+M|=d$, we need either of the two equivalents
in Eq.~\eqref{eq:s-antisymm-boundary}.
When $\widetilde{C}=1$, as $s$ can take only one value, this 
antisymmetrization consistently
reduces to an electrostatic repulsion at distance $|j_1-j_2|=d$ (and also 
$|j_1-j_2+M|=d$).

\subsection{Counting Rule for Degeneracy and Momenta}\label{sec:counting-rule}

Following the above constraints, we can enumerate all the zero modes of the 
truncated Hamiltonian for a given system size and a given number of particles, 
in the form
\begin{equation}\label{eq:zero-modes}
\mathcal{A}\big[\,\psi^\dagger_{j_1,s_1}\psi^\dagger_{j_2,s_2}
\psi^\dagger_{j_3,s_3}\psi^\dagger_{j_4,s_4}\cdots\,\big]
\,|\emptyset\rangle,
\end{equation}
where $\mathcal{A}$ antisymmetrizes the $s$ indices as follows.
As noted earlier, for any pair of particles $a$ and $b$ in a zero mode, we 
must have $|j_a-j_b|\ge d$, and when the equality holds, we need to carry out 
antisymmetrization over the $s$ indices $(s_a,s_b)$.
Obviously, if we have $j_1-j_2=d$ and $j_2-j_3=d$, then we need to 
antisymmetrize over $(s_1,s_2,s_3)$.
More generally, if we have a cluster of $m$ consecutive particles satisfying 
$j_a-j_{a+1}=d$, we need a full antisymmetrization over all the $s$ indices of 
these $m$ particles.

The last remaining step is to group these zero modes according by the total 
Bloch momentum and count the degeneracy per momentum sector.
The resulting degeneracy is linked by the Bloch mapping~\cite{Wu13:Bloch} to 
the degeneracy of FCI ground states per lattice momentum sector.
This largely follows the same procedure as detailed in 
Ref.~\onlinecite{Bernevig12:Counting}.
We represent by lowercase $k_\alpha$ the Bloch
momenta of individual particles in the $\alpha=x,y$ direction,
and by uppercase
\begin{equation}
K_\alpha=\sum k_\alpha\text{ mod }N_\alpha
\end{equation}
the total Bloch momentum
of the many-body system (the summation is over particles).
We denote by $\widetilde{T}^\mathrm{cm}_\alpha$ the center-of-mass 
color-entangled magnetic translations, i.e. applying $\widetilde{T}_\alpha$ 
simultaneously on all the particles.
Then, the total Bloch momentum $K_\alpha$ can be read off from the eigenvalue
of $\widetilde{T}^\mathrm{cm}_\alpha$,
\begin{equation}
\widetilde{T}^\mathrm{cm}_\alpha=e^{-i2\pi K_\alpha/N_\alpha}.
\end{equation}
The action 
of $\widetilde{T}^\mathrm{cm}_\alpha$ on the zero modes in 
Eq.~\eqref{eq:zero-modes} is spelled out in 
Eq.~\eqref{eq:Wannier-translations-relabeled}.

There are four points to make here.
\emph{First}, the zero modes in the form of Eq.~\eqref{eq:zero-modes} are 
automatically eigenstates of $\widetilde{T}^\mathrm{cm}_y$.
Evidently each term in the antisymmetrization $\mathcal{A}$ individually is an 
eigenstate of $\widetilde{T}^\mathrm{cm}_y$.
Moreover, the eigenvalues have to be the same for all those terms.
This follows from the linearity of Eq.~\eqref{eq:j-s-def}: to find the total 
$\sum k_y$ of all particles, we only need to know the total $\sum j$ and 
$\sum s$; the actual association of between $j$ and $s$ does not matter.
\emph{Second}, under the action of $\widetilde{T}^\mathrm{cm}_x$, the zero modes in 
Eq.~\eqref{eq:zero-modes} form closed orbits.
This follows from the fact that $\widetilde{T}^\mathrm{cm}_x$ commutes 
with the (truncated) pseudopotential Hamiltonian, and thus preserves its null 
space. More directly, one can easily verify that the constraints on the 
zero modes described in Secs.~\ref{sec:zero-modes-d1} 
and~\ref{sec:zero-modes-general-d} are invariant under the action of 
$\widetilde{T}^\mathrm{cm}_x$ (namely $X\rightarrow X+1$, or
$|\{j,s\}\rangle\rightarrow |\{j+N_y/\widetilde{C},s+1\}\rangle$), and that 
the action of $\widetilde{T}^\mathrm{cm}_x$ always brings one zero mode in the 
form of Eq.~\eqref{eq:zero-modes} to another zero mode in the same form.
\emph{Third}, each action of $\widetilde{T}^\mathrm{cm}_x$ along the orbit is 
associated with a sign, since a term in the antisymmetrization $\mathcal{A}$ 
in Eq.~\eqref{eq:zero-modes} may be brought to a term with the opposite sign.
\footnote{For the case of fermions, there also is a statistical sign, as noted in 
Ref.~\onlinecite{Bernevig12:Counting}.}
\emph{Fourth}, all the zero modes in an orbit under 
$\widetilde{T}^\mathrm{cm}_x$ share the same eigenvalue under 
$\widetilde{T}^\mathrm{cm}_y$. This is a direct consequence of 
$[\widetilde{T}^\mathrm{cm}_x,\widetilde{T}^\mathrm{cm}_y]=0$.

For each zero mode in the form of Eq.~\eqref{eq:zero-modes}, we can directly 
compute the total $K_y$ momentum by just looking at a single term in the 
antisymmetrization $\mathcal{A}$.
We can group together the zero modes by the value of $K_y=\sum k_y$ mod $N_y$.
Then, within each group, we successively apply $\widetilde{T}^\mathrm{cm}_x$ 
on each zero mode and further break them into disjoint orbits.
Consider an orbit consisting of $n$ zero modes 
$|0\rangle,|1\rangle,\ldots,|n-1\rangle$ of the form in Eq.~\eqref{eq:zero-modes}.
They are linked together by
\begin{equation}\label{eq:Tcmx-orbit}
\widetilde{T}^\mathrm{cm}_x|r\rangle=g_r\,|r+1\text{ mod }n\rangle,
\quad r\in[0~..~n),
\end{equation}
with $g_r=\pm 1$ determined from the action of $\widetilde{T}_x^\mathrm{cm}$ 
on the antisymmetrization in Eq.~\eqref{eq:zero-modes}.
The $n$ eigenstates of $\widetilde{T}^\mathrm{cm}_x$ are linear recombinations 
of these $n$ states in the form of Fourier sums.
Without actually writing down the linear recombinations, we can directly obtain 
the eigenvalues.
By successively applying the above equation, we find
\begin{equation}
[\widetilde{T}^\mathrm{cm}_x]^n\,|r\rangle=g\,|r\rangle,
\end{equation}
with $g=\prod_{r'}^n g_{r'}$.
This fixes the $n$ eigenvalues of $\widetilde{T}^\mathrm{cm}_x$ to be the $n$ 
distinct $n$-th roots of $g$.
If $g=1$, the total $K_x$ momenta of the zero modes are
\begin{equation}
K_x = k \frac{N_x}{n}\text{ mod }N_x,
\quad k\in[0~..~n),
\end{equation}
whereas if $g=-1$, they are
\begin{equation}
K_x = k \frac{N_x}{n} + \frac{N_x}{2}\text{ mod }N_x,
\quad k\in[0~..~n).
\end{equation}
The numbers on the right hand side of the above equation are guaranteed to be 
integers:
Since $[\widetilde{T}^\mathrm{cm}_x]^{N_x}$ is the identity operator per the 
color-entangled boundary condition 
[Eq.~\eqref{eq:color-entangled-boundaries}], we must have $N_x/n\in\mathbb{Z}$, 
and also $g^{N_x/n}=1$.

Our end goal is an analytic algorithm to obtain the degeneracy of the zero modes 
in each Bloch momentum sector.
This request is more modest than to find the actual expression of the zero 
modes in each sector, and the above procedure can be further simplified.
For example, we do not need to actually write down the zero modes as in 
Eq.~\eqref{eq:zero-modes}.
We only need to keep track of the structure of clusters of consecutive particles 
with $j_a-j_{a+1}=d$, as noted below Eq.~\eqref{eq:zero-modes}, and the set of 
$s$ indices in each cluster.
An open-source reference implementation can be found at
\url{http://fractionalized.github.io}.
We have tested our algorithm extensively against the total Bloch momenta of 
the actual ground states obtained from numerical diagonalization for various 
system sizes, and found perfect agreement across all cases.

\subsection{A Simple Example}\label{sec:counting-example}

\begin{figure}[tb]
\centering
\includegraphics[]{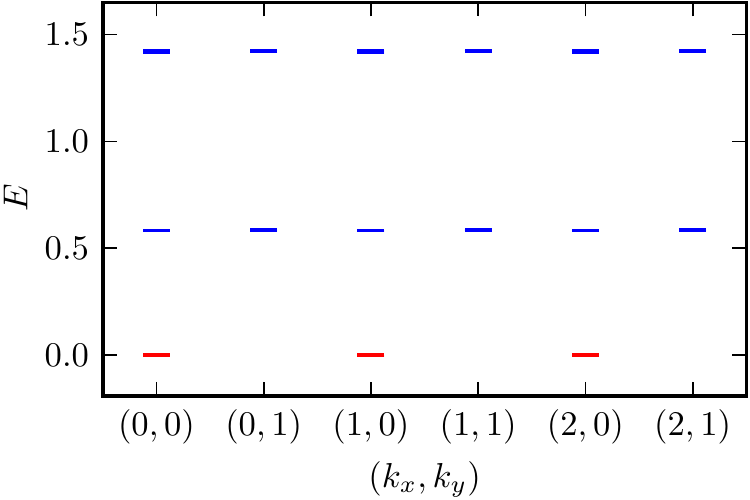}%
\caption{\label{fig:example-spectrum}
Energy spectrum of the pseudopotential Hamiltonian of 2 bosons on
a $N_x\times N_y=3\times 2$ lattice with Chern number $C=2$.
The three degenerate ground states at zero energy are marked in red.
}
\end{figure}

To see the above counting rule in action, we consider a simple example,
2 bosons on a $N_x\times N_y=3\times 2$ lattice with Chern number $C=2$.
From numerical diagonalization of the pseudopotential Hamiltonian
(see Fig.~\ref{fig:example-spectrum}),
we find 3-fold degenerate ground states with total Bloch momenta 
\begin{equation}\label{eq:example-correct-K}
(K_x,K_y)=(0,0),(1,0),(2,0)\text{ mod }(3,2).
\end{equation}
We note that the spinless counting rule~\cite{Bernevig12:Counting}
gives the wrong result $K_y=1$ mod $2$ when naively applied to this system.
We now show how our new procedure produces the correct momenta.

From $(N_x,N_y)=(3,2)$ and $C=2$, we compute 
$\widetilde{C}=\mathrm{GCD}(C,N_y)=2$, $d=C/\widetilde{C}=1$, 
$M=N_xN_y/\widetilde{C}=3$.
Equation~\eqref{eq:j-s-def} reduces to
\begin{equation}\label{eq:j-s-def-example}
\begin{aligned}
j&=X+k_y,\\
s&=X\text{ mod }2.
\end{aligned}
\end{equation}
We denote $s=0$ by $\downarrow$ and $s=1$ by $\uparrow$.
To facilitate two-way lookup of the mapping $(X,k_y)\leftrightarrow(j,s)$, we 
can make a table
\begin{equation}\label{eq:Xky-js-lookup}
\begin{tabular}{C{2em} | C{2em} || C{2em} | C{2em}}
$X$ & $k_y$ & $j$ & $s$ \\\hline\hline
0 & 0 & 0 & $\downarrow$ \\\hline
0 & 1 & 1 & $\downarrow$ \\\hline
1 & 0 & 1 & $\uparrow$ \\\hline
1 & 1 & 2 & $\uparrow$ \\\hline
2 & 0 & 2 & $\downarrow$ \\\hline
2 & 1 & 0 & $\uparrow$
\end{tabular}
\end{equation}
The last line in the above table deserves special attention.
From Eq.~\eqref{eq:j-s-def-example}, for $(X,k_y)=(2,1)$ we obtain 
$(j,s)={(3,\downarrow)}$. However,
due to the quasi-periodicity condition in $j$,
[Eq.~\eqref{eq:Wannier-periodicity-relabeled}],
this is equivalent to $(j,s)={(0,\uparrow)}$.

We enumerate all the two-boson zero modes of the truncated pseudopotential 
Hamiltonian [Eq.~\eqref{eq:H-truncated-boson}] in the form of 
Eq.~\eqref{eq:zero-modes}.
Applying the constraint $|j_1-j_2|\ge 1$ across the quasi-periodic boundary of 
$j$, we find only three possibilities
\begin{equation}
(j_1,j_2)=(0,1),(1,2),(0,2).
\end{equation}
All of them satisfy either $|j_1-j_2|=d$ (first two) or $|j_1-j_2+M|=d$ (last 
one), and are thus 
subject to full antisymmetrization of the $s$ indices $(s_1,s_2)$.
Since there are only two allowed values of $s$, we can already see that there 
are only 3 zero modes in the form of Eq.~\eqref{eq:zero-modes}.
We now go through them one by one.
First, consider $(j_1,j_2)=(0,1)$. Using Eq.~\eqref{eq:s-antisymm}, we find that the 
only possible $(s_1,s_2)$ antisymmetrization is
\begin{equation}
|0,1\rangle\!\rangle
\equiv
\left(
\psi^\dagger_{0,\downarrow}\psi^\dagger_{1,\uparrow}
-\psi^\dagger_{0,\uparrow}\psi^\dagger_{1,\downarrow}
\right)\,
\big|\emptyset\big\rangle.
\end{equation}
Here the double bracket $|\cdot,\cdot\rangle\!\rangle$
distinguishes the many-body zero mode from the one-body basis state $|j,s\rangle$,
and the subscript of the creation operator $\psi^\dagger$ denotes $(j,s)$.
Similarly, for $(j_1,j_2)=(1,2)$, we find
\begin{equation}
|1,2\rangle\!\rangle
\equiv
\left(
\psi^\dagger_{1,\downarrow}\psi^\dagger_{2,\uparrow}
-\psi^\dagger_{1,\uparrow}\psi^\dagger_{2,\downarrow}
\right)\,
\big|\emptyset\big\rangle.
\end{equation}
The case of $(j_1,j_2)=(0,2)$ satisfies $|j_1-j_2+M|=d$ rather than 
$|j_1-j_2|=d$. So we use Eq.~\eqref{eq:s-antisymm-boundary} rather than 
Eq.~\eqref{eq:s-antisymm}, and find
\begin{equation}\label{eq:example-antisymm-boundary}
\begin{aligned}
|0,2\rangle\!\rangle
&\equiv\left(
\psi^\dagger_{3,\downarrow}\psi^\dagger_{2,\uparrow}
-\psi^\dagger_{3,\uparrow}\psi^\dagger_{2,\downarrow}
\right)\,
\big|\emptyset\big\rangle\\
&=\left(
\psi^\dagger_{0,\uparrow}\psi^\dagger_{2,\uparrow}
-\psi^\dagger_{0,\downarrow}\psi^\dagger_{2,\downarrow}
\right)\,
\big|\emptyset\big\rangle.
\end{aligned}
\end{equation}
Notice that after we bring the $j$ indices back to $[0~..~M)$ using
Eq.~\eqref{eq:Wannier-quasiperiodicity-relabeled}, the $s$ indices on the 
second line are \emph{not} in an explicit antisymmetrized form.
This manifests the core difference of our problem from the usual FQH:
When the lattice size is 
incommensurate with the Chern number, we cannot consistently distinguish the 
$C$ families of Wannier states, since the flow of Wannier centers are 
entangled on the quasi-perioidic boundary.

Using the lookup table in Eq.~\eqref{eq:Xky-js-lookup},
we find that the total $K_y$ momenta of the three 
zero modes are all equal to $0$ mod $2$, consistent with 
Eq.~\eqref{eq:example-correct-K}.
To compute the $K_x$ momentum,
we need to find out the action of the center-of-mass 
translation $\widetilde{T}_x^\mathrm{cm}$ on these states.
For our example, Equation~\eqref{eq:Wannier-translations-relabeled} reduces to
\begin{equation}
\widetilde{T}_x|j,s\rangle=|j+1,s+1\rangle.
\end{equation}
We thus find the representation of $\widetilde{T}_x^\mathrm{cm}$ on the zero modes:
\begin{equation}
\begin{aligned}
\widetilde{T}_x^\mathrm{cm}|0,1\rangle\!\rangle
&=\left(
\psi^\dagger_{1,\uparrow}\psi^\dagger_{2,\downarrow}
-\psi^\dagger_{1,\downarrow}\psi^\dagger_{2,\uparrow}
\right)
\big|\emptyset\big\rangle
=-|1,2\rangle\!\rangle,\\
\widetilde{T}_x^\mathrm{cm}|1,2\rangle\!\rangle
&=\left(
\psi^\dagger_{2,\uparrow}\psi^\dagger_{3,\downarrow}
-\psi^\dagger_{2,\downarrow}\psi^\dagger_{3,\uparrow}
\right)
\big|\emptyset\big\rangle
=|0,2\rangle\!\rangle,\\
\widetilde{T}_x^\mathrm{cm}|0,2\rangle\!\rangle
&=\left(
\psi^\dagger_{1,\downarrow}\psi^\dagger_{3,\downarrow}
-\psi^\dagger_{1,\uparrow}\psi^\dagger_{3,\uparrow}
\right)
\big|\emptyset\big\rangle\\
&=\left(
\psi^\dagger_{1,\downarrow}\psi^\dagger_{0,\uparrow}
-\psi^\dagger_{1,\uparrow}\psi^\dagger_{0,\downarrow}
\right)
\big|\emptyset\big\rangle
=-|0,1\rangle\!\rangle.
\end{aligned}
\end{equation}
Notice that we can evaluate $\widetilde{T}_x^\mathrm{cm}|0,2\rangle\!\rangle$
using either line in Eq.~\eqref{eq:example-antisymm-boundary};
the results are guaranteed to be the same by the consistency between
Eqs.~\eqref{eq:Wannier-translations-relabeled}
and~\eqref{eq:Wannier-periodicity-relabeled}.

The three zero modes thus form a single orbit under the successive action of 
$\widetilde{T}_x^\mathrm{cm}$.
They can be recombined to form eigenstates of $\widetilde{T}_x^\mathrm{cm}$.
To find the total $K_x$ momenta of the recombined states, we can either follow 
the procedure detailed in the last subsection,
or we can brute-force diagonalize 
$\widetilde{T}_x^\mathrm{cm}$.
The representation matrix of $\widetilde{T}_x^\mathrm{cm}$ over 
the three zero modes reads
\begin{equation}
\left(
\begin{array}{ccc}
0 & -1 & 0 \\
0 & 0 & 1 \\
-1 & 0 & 0
\end{array}
\right).
\end{equation}
From its eigenvalues $\{1,e^{i2\pi/3},e^{-i2\pi/3}\}$,
we find the total $K_x$ momenta of the three recombined zero modes to be 
$0,1,2$ mod $3$.
In summary, we reproduce the correct total Bloch momenta in 
Eq.~\eqref{eq:example-correct-K}.

\begin{figure}[tb]
\centering
\includegraphics[]{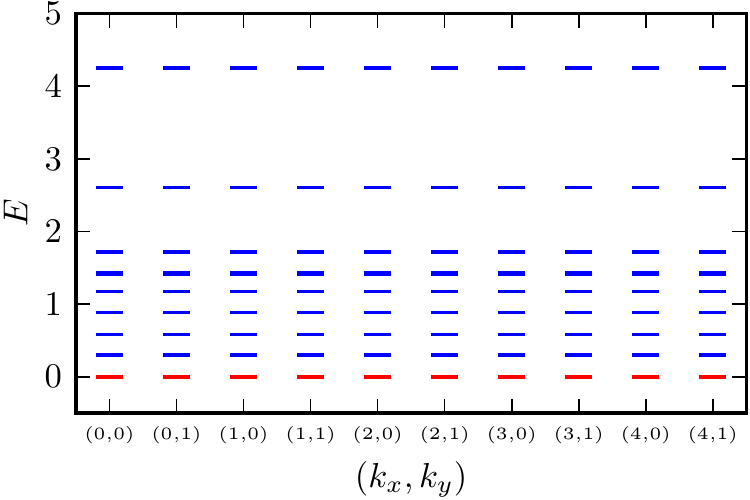}%
\caption{\label{fig:example-spectrum-quasiholes}
Energy spectrum of the pseudopotential Hamiltonian of 3 bosons on
a $N_x\times N_y=5\times 2$ lattice with Chern number $C=2$.
The 10 degenerate ground states at zero energy are marked in red.
}
\end{figure}

\subsection{An Example with Quasiholes}\label{sec:counting-example-qh}

Next, we consider a slightly more complicated example with quasiholes.
For a system of 3 bosons with $C=2$, the densest zero modes of our
pseudopotential Hamiltonian occur at filling $\nu=1/(C+1)=1/3$, i.e. 
3 bosons in $9/2$ fluxes.
The fractional flux is possible thanks to the color-entangled boundary
conditions in Eq.~\eqref{eq:color-entangled-boundaries}.
We now add $\frac{1}{2}$ flux to each color component
and consider $N_x\times N_y=5\times 2$ and $N_\phi=9/2+1/2=N_xN_y/C=5$.
This leads to a set of 10-fold degenerate quasihole states at zero energy,
with one mode in each momentum sector
$(K_x,K_y)\in[0~..~N_x)\times [0~..~N_y)$.
This can be seen in the numerical diagonalization results in
Fig.~\ref{fig:example-spectrum-quasiholes}.

We now show how to obtain this counting using our algorithm.
The basic procedure is the same as the previous example.
We first compute
$\widetilde{C}=\mathrm{GCD}(C,N_y)=2$, $d=C/\widetilde{C}=1$,
and $M=N_xN_y/\widetilde{C}=5$.
Equation~\eqref{eq:j-s-def} again reduces to Eq.~\eqref{eq:j-s-def-example},
and we have two allowed values of $s$ ($0$ and $1$),
denoted by $\downarrow$ and $\uparrow$.
Then we can build the $(X,k_y)\leftrightarrow(j,s)$ lookup table,
\begin{equation}\label{eq:Xky-js-lookup-qh}
\begin{tabular}{C{2em} | C{2em} || C{2em} | C{2em}}
$X$ & $k_y$ & $j$ & $s$ \\\hline\hline
0 & 0 & 0 & $\downarrow$ \\\hline
0 & 1 & 1 & $\downarrow$ \\\hline
1 & 0 & 1 & $\uparrow$ \\\hline
1 & 1 & 2 & $\uparrow$ \\\hline
2 & 0 & 2 & $\downarrow$ \\\hline
2 & 1 & 3 & $\downarrow$ \\\hline
3 & 0 & 3 & $\uparrow$ \\\hline
3 & 1 & 4 & $\uparrow$ \\\hline
4 & 0 & 4 & $\downarrow$ \\\hline
4 & 1 & 0 & $\uparrow$
\end{tabular}
\end{equation}
Again, the last line in the table has a flipped $s$ index due to the 
quasi-periodic boundary condition in $j$
[Eq.~\eqref{eq:Wannier-periodicity-relabeled}].

Compared with the previous example, the enumeration of the zero modes
in the form of Eq.~\eqref{eq:zero-modes} has an extra complication.
Let us first apply the $|j_a-j_b|\ge 1$ rule between each pair of bosons.
We find two groups of allowed $(j_1,j_2,j_3)$ configurations,
\begin{equation}\label{eq:cluster-3}
(\underline{0,1,2}),(\underline{1,2,3}),(\underline{2,3,4}),
(\underline{3,4,0}),(\underline{4,0,1}),
\end{equation}
and
\begin{equation}\label{eq:cluster-2}
(\underline{0,1},3),(\underline{1,2},4),(\underline{2,3},0),
(\underline{3,4},1),(\underline{4,0},2).
\end{equation}
Here we have underlined each cluster of bosons linked together
by $|j_a-j_b|=d=1$ or $|j_a-j_b+M|=d=1$.
Then, we need to antisymmetrize the $s$ indices of the bosons
in the same cluster.
This kills the five configurations in the first group
[Eq.~\eqref{eq:cluster-3}]:
the three bosons in the same cluster must take different values
of $s$ under antisymmetrization,
but there are only two possible values of $s$ ($\uparrow$ and $\downarrow$).
We are left with the five $j$ configurations in Eq.~\eqref{eq:cluster-2}.
In each configuration, the two clustered bosons have antisymmetrized $s$ 
indices $\uparrow\downarrow-\downarrow\uparrow$, while the third boson
can take either $\uparrow$ or $\downarrow$.
For example, for $(j_1,j_2,j_3)=(0,1,3)$, we have a pair of zero modes
\begin{equation}
\begin{aligned}
&\left(
\psi^\dagger_{0,\downarrow}\psi^\dagger_{1,\uparrow}
-\psi^\dagger_{0,\uparrow}\psi^\dagger_{1,\downarrow}
\right)
\psi^\dagger_{3,\uparrow}\,
\big|\emptyset\big\rangle,\\
&\left(
\psi^\dagger_{0,\downarrow}\psi^\dagger_{1,\uparrow}
-\psi^\dagger_{0,\uparrow}\psi^\dagger_{1,\downarrow}
\right)
\psi^\dagger_{3,\downarrow}\,
\big|\emptyset\big\rangle.
\end{aligned}
\end{equation}
We can similarly write down the other 8 zero modes.
This gives the correct 10-fold degeneracy.
Using the lookup table in Eq.~\eqref{eq:Xky-js-lookup-qh},
we can compute the $K_y$ lattice momentum
for each zero mode and construct the representation matrix
of the color-entangled center-of-mass translation operator 
$\widetilde{T}_x^\mathrm{cm}$ in exactly the same manner as in the previous 
example. This reproduces the correct degeneracy in each momentum sector.
We leave details of this last step for the interested readers.

\section{Conclusion}

In this paper we have studied the pseudopotential model Hamiltonian for FCI 
with an arbitrary Chern number.
We establish a one-body mapping between a Chern band with Chern number $C$, 
and a $C$-component LLL with specially engineered boundary conditions.
The new boundary conditions lead to an alternative set of pseudopotential 
Hamiltonians, and the corresponding zero modes define new model wave functions.
By taking the thin-torus limit and keeping only the leading density-density 
and pair hopping terms, we are able to analytically solve the pseudopotential 
Hamiltonian and obtain its zero modes.
By analyzing the representation of the center-of-mass translation operators, 
we derive an algorithm to directly compute the total Bloch momenta of the 
degenerate zero modes.
As we showed in our last paper~\cite{Wu13:Bloch}, our pseudopotential 
Hamiltonian is adiabatically connected to the lattice FCI Hamiltonian, and the 
its zero modes serve as good trial wave functions for the FCI ground states.
In particular, there is a 1-to-1 correspondence between the trial wave 
function and the FCI ground state in each momentum sector.
Therefore, our counting algorithm can be used to obtain the total lattice 
momenta of the FCI ground states (including quasiholes) without diagonalizing 
the FCI Hamiltonian, for Abelian FCI states at filling $\nu=1/(C+1)$.

\section*{Acknowledgments}

We wish to thank C.~Fang, B.~Estienne, A.~Sterdyniak, C.~Laumann, and N.Y.~Yao 
for useful discussions.
BAB and NR were supported by NSF CAREER DMR-095242, ONR-N00014-11-1-0635, 
ARMY-245-6778, MURI-130-6082, Packard Foundation, and Keck grant.
YLW was supported by NSF CAREER DMR-095242.

\appendix

\section{Hybrid Wannier States under Color-Entangled Magnetic Translations}
\label{sec:Wannier-translations}

In this Appendix we prove Eq.~\eqref{eq:Wannier-translations}, the 
representation of $\widetilde{T}_x$ and $\widetilde{T}_y$ in the hybrid 
Wannier basis $|X,k_y\rangle$.

In Landau gauge $\mathbf{A}=Bx\hat{y}$, the magnetic translation operators 
$T_x$ and $T_y$ defined in Eq.~\eqref{eq:TxTy} have the real-space representation
\begin{equation}
\begin{aligned}
T_x&=e^{i2\pi \frac{N_\phi}{N_x}\frac{y}{L_y}}e^{-\frac{L_x}{N_x}\partial_x},\\
T_y&=e^{-\frac{L_y}{N_y}\partial_y}.
\end{aligned}
\end{equation}
Acting on a trial state $|\psi\rangle$, they transform
the real-space wave function 
$\langle x,y|\psi\rangle$ by
\begin{equation}
\begin{aligned}
\langle x,y|T_x|\psi\rangle
&=e^{i2\pi \frac{N_\phi}{N_x}\frac{y}{L_y}}
\langle x-L_x/N_x,y|\psi\rangle,\\
\langle x,y|T_y|\psi\rangle
&=\langle x,y-L_y/N_y|\psi\rangle.
\end{aligned}
\end{equation}
Plugging these into the Landau-gauge definition of $|X,k_y\rangle$ in 
Eq.~\eqref{eq:Wannier-wf} and using Eq.~\eqref{eq:Nf-def}, we find
\begin{align}\label{eq:Wannier-TxTy}
\langle x,y,\sigma|T_x|X,k_y\rangle
&=\langle x,y,\sigma+1|X+1,k_y\rangle,\\
\nonumber
\langle x,y,\sigma|T_y|X,k_y\rangle
&=e^{-i2\pi (k_y/N_y+\sigma/C)}
\langle x,y,\sigma|X,k_y\rangle.
\end{align}
Since the clock-and-shift operators $Q,P$ defined in 
Eq.~\eqref{eq:clock-and-shift} are unitary, we have
\begin{equation}\label{eq:Wannier-PQ}
\begin{aligned}
\langle\sigma|P&=\langle{\sigma-1}|,&
\langle\sigma|Q&=e^{i2\pi\sigma/C}\langle\sigma|.
\end{aligned}
\end{equation}
Putting Eqs.~\eqref{eq:Wannier-TxTy} and~\eqref{eq:Wannier-PQ} together, we 
find the action of $\widetilde{T}_x=T_xP$ and $\widetilde{T}_y=T_yQ$ to be 
particularly simple,
\begin{equation}
\begin{aligned}
\langle x,y,\sigma|\widetilde{T}_x|X,k_y\rangle
&=\langle x,y,\sigma|X+1,k_y\rangle,\\
\langle x,y,\sigma|\widetilde{T}_y|X,k_y\rangle
&=e^{-i2\pi k_y/N_y}
\langle x,y,\sigma|X,k_y\rangle.
\end{aligned}
\end{equation}
This proves Eq.~\eqref{eq:Wannier-translations}.

\section{Projected Density in Bloch Basis}\label{sec:density-Bloch}

In this Appendix we prove Eq.~\eqref{eq:density-Bloch}, the expansion of the 
projected density operator in the Bloch basis,
proof which, due to lack of space, was not included in 
Ref.~\onlinecite{Wu13:Bloch}.

We first derive a simpler form for the Bloch wave function
$\phi_\mathbf{k}(\mathbf{r},\sigma)=\langle\mathbf{r},\sigma|\mathbf{k}\rangle$.
When we plug Eq.~\eqref{eq:Wannier-wf} into Eq.~\eqref{eq:Bloch-def},
we have a double sum over $X,m$. However, notice that
$(X,m)$ in the double sum can always be combined into $X+mN_x$, thanks to
$XN_y/C+mN_\phi=(X+mN_x)N_y/C$ and 
$e^{i2\pi Xk_x/N_x}=e^{i2\pi (X+mN_x)k_x/N_x}$ (recall that $N_\phi=N_xN_y/C$).
This enables us to merge the double sum into a single sum of $X+mN_x$ over 
$\mathbb{Z}$. The Kronecker-$\delta$ enforcing $\sigma=X+mN_x$ mod $C$ 
suggests we split $X+mN_x\rightarrow nC+\sigma$ with $n$ summed over 
$\mathbb{Z}$.
This leads to the final form of the Bloch wave function,
\begin{widetext}
\begin{multline}
\langle x,y,\sigma|\mathbf{k}\rangle
=\frac{1}{(\sqrt{\pi}N_xL_yl_B)^{1/2}}
\sum_n^{\mathbb{Z}}
e^{i2\pi (nC+\sigma)k_x/N_x}\\
\exp\left\{
i2\pi\left(k_y+nN_y+\frac{\sigma}{C}N_y\right)\frac{y}{L_y}
-\frac{1}{2}\left[
\frac{x}{l_B}-\frac{2\pi l_B}{L_y}\left(k_y+nN_y+\frac{\sigma}{C}N_y\right)
\right]^2
\right\}.
\end{multline}
This wave function indeed depends only on $\sigma$ mod $C$ (by a re-shift in 
the dummy variable $n$), and it has the quasi-periodicity in $k_y$ as in 
Eq.~\eqref{eq:Bloch-periodicity}.
We now plug this into $\rho_{\mathbf{q},\sigma}$ defined in 
Eq.~\eqref{eq:rho-q-integral}.
\begin{multline}
\rho_{\mathbf{q},\sigma}
=\frac{1}{\sqrt{\pi}N_xL_yl_B}
\sum_{\mathbf{k}_1}^\mathrm{BZ}\sum_{\mathbf{k}_2}^\mathrm{BZ}
|\mathbf{k}_1\rangle\langle\mathbf{k}_2|
\sum_{n_1}^{\mathbb{Z}}\sum_{n_2}^{\mathbb{Z}}
e^{-i2\pi(n_1 C+\sigma)k_{1x}/N_x}
e^{i2\pi(n_2 C+\sigma)k_{2x}/N_x}\\
\left[
\int_0^{L_x}\mathrm{d}x\,
e^{-i2\pi q_x x/L_x}
\exp\left\{
-\frac{1}{2}\left[\frac{x}{l_B}
-\frac{2\pi l_B}{L_y}\left(k_{1y}+\frac{\sigma}{C}N_y+n_1N_y\right)\right]^2
-\frac{1}{2}\left[\frac{x}{l_B}
-\frac{2\pi l_B}{L_y}\left(k_{2y}+\frac{\sigma}{C}N_y+n_2N_y\right)\right]^2
\right\}
\right]\\
\left[
\int_0^{L_y}\mathrm{d}y\,
e^{-i2\pi q_y y/L_y-i2\pi(k_{1y}+n_1 N_y-k_{2y}-n_2 N_y)y/L_y}
\right].
\end{multline}
We first finish the $\int\mathrm{d}y$ integral on the last line,
\begin{equation}
\int\mathrm{d}y\,e^{-i2\pi\cdots}
=L_y\,\delta_{k_{2y}+n_2N_y,\;k_{1y}+n_1N_y+q_y}.
\end{equation}
Notice that the summations of $\mathbf{k}_1$ and $\mathbf{k}_2$ over BZ in the 
above equation are independent.
To accommodate the Kronecker-$\delta$ in the above equation, we set the 
summation of $k_{1y}$ over $[0~..~N_y)$, and the summation of $k_{2y}$ over 
$[q_y~..~N_y+q_y)$.
Then, the Kronecker-$\delta$ above can be decomposed into two separate 
Kronecker-$\delta$'s, enforcing
\begin{align}
\begin{aligned}
k_{2y}&=k_{1y}+q_y,\\
n_1&=n_2.
\end{aligned}
\end{align}
And we have
\begin{multline}
\rho_{\mathbf{q},\sigma}
=\frac{1}{\sqrt{\pi}N_xl_B}
\sum_{k_{1x}}^{N_x}\sum_{k_{2x}}^{N_x}
\sum_{k_y}^{N_y}
|k_{1x},k_y\rangle\langle k_{2x},{k_y+q_y}|
\sum_{n}^{\mathbb{Z}}
e^{i2\pi(n C+\sigma)(k_{2x}-k_{1x})/N_x}\\
\left[
\int_0^{L_x}\mathrm{d}x\,
e^{-i2\pi q_x x/L_x}
\exp\left\{
-\frac{1}{2}\left[\frac{x}{l_B}
-\frac{2\pi l_B}{L_y}\left(k_y+\frac{\sigma}{C}N_y+nN_y\right)\right]^2
-\frac{1}{2}\left[\frac{x}{l_B}
-\frac{2\pi l_B}{L_y}\left(k_y+q_y+\frac{\sigma}{C}N_y+nN_y\right)\right]^2
\right\}
\right].
\end{multline}
It is easy to check that $\rho_{\mathbf{q},\sigma}$ is indeed invariant
under a shift of the dummy variable $k_y\rightarrow k_y+N_y$.
We now tackle the $\int\mathrm{d}x$ integral in the bracket.
We can collect terms and complete the square in the exponential.
After some trivial but tedious algebra, the integrand becomes
\begin{equation}
e^{-\mathbf{q}^2l_B^2/4}\,
e^{-i2\pi q_x(k_y+q_y/2+\sigma N_y/C+nN_y)/N_\phi}
\exp\left\{
-\left[\frac{x}{l_B}
-\frac{2\pi l_B}{L_y}
\left(k_y+\frac{\sigma}{C}N_y+nN_y
+\frac{1}{2}\left(q_y-i\frac{L_y}{L_x}q_x\right)
\right)\right]^2
\right\}.
\end{equation}
Here we have used $2\pi l_B^2N_\phi=L_xL_y$ [Eq.~\eqref{eq:Nf-def}], and
\begin{equation}
\mathbf{q}^2l_B^2
=\frac{2\pi}{N_\phi}\left(\frac{L_y}{L_x}q_x^2+\frac{L_x}{L_y}q_y^2\right).
\end{equation}
The projected density can thus be written as
\begin{multline}
\rho_{\mathbf{q},\sigma}
=\frac{1}{\sqrt{\pi}N_xl_B}
e^{-\mathbf{q}^2l_B^2/4}
\sum_{k_{1x}}^{N_x}\sum_{k_{2x}}^{N_x}
\sum_{k_y}^{N_y}
|k_{1x},k_y\rangle\langle k_{2x},{k_y+q_y}|\,
e^{-i2\pi q_x(k_y+q_y/2)/N_\phi}\\
\sum_{n}^{\mathbb{Z}}
e^{i2\pi(n C+\sigma)(k_{2x}-k_{1x}-q_x)/N_x}\
\int_0^{L_x}\mathrm{d}x
\exp\left\{
-\left[\frac{x}{l_B}
-\frac{2\pi l_B}{L_y}
\left(k_y+\frac{\sigma}{C}N_y+nN_y
+\frac{1}{2}\left(q_y-i\frac{L_y}{L_x}q_x\right)
\right)\right]^2
\right\}.
\end{multline}
\end{widetext}
Notice that
\begin{equation}
\frac{x}{l_B}-\frac{2\pi l_B}{L_y}nN_y=\frac{x-nN_yL_x/N_\phi}{l_B};
\end{equation}
we can shift the integration interval to
\begin{equation}
\int_{n\frac{N_y}{N_\phi}L_x}^{L_x+n\frac{N_y}{N_\phi}L_x}
\mathrm{d}x.
\end{equation}
This moves the dependence on $n$ from the integrand
to the integration limits (and also the exponential prefactor
$e^{i2\pi(nC+\sigma)(k_{2x}-k_{1x}-q_x)/N_x}$).

We want to sew together the integrals for all $n$ so that we can finish the 
Gaussian integral, but the integration intervals for different $n$ are 
overlapping and cannot be joined head to tail in general, unless $N_x$ is 
divisible by $C$.
However, recall that (to have symmetries $P,Q$) we restrict the interacting 
Hamiltonian to be color-neutral, so we are interested only in 
$\rho_\mathbf{q}=\sum_\sigma^C\rho_{\mathbf{q},\sigma}$. The color sum saves us.
Notice that the dependence on $(\sigma,n)$ is all through the 
combination $nC+\sigma$. We can merge the two sums over $\sigma$ and $n$ into 
a single sum over integers, $nC+\sigma\rightarrow m$:
\begin{widetext}
\begin{multline}
\rho_\mathbf{q}
=\frac{1}{\sqrt{\pi}N_xl_B}
e^{-\mathbf{q}^2l_B^2/4}
\sum_{k_{1x}}^{N_x}\sum_{k_{2x}}^{N_x}
\sum_{k_y}^{N_y}
|k_{1x},k_y\rangle\langle k_{2x},{k_y+q_y}|\,
e^{-i2\pi q_x(k_y+q_y/2)/N_\phi}\\
\sum_{n}^{\mathbb{Z}}
e^{i2\pi n(k_{2x}-k_{1x}-q_x)/N_x}\
\int_0^{L_x}\mathrm{d}x
\exp\left\{
-\left[\frac{x}{l_B}
-\frac{2\pi l_B}{L_y}
\left(k_y+\frac{n}{C}N_y
+\frac{1}{2}\left(q_y-i\frac{L_y}{L_x}q_x\right)
\right)\right]^2
\right\}.
\end{multline}
Notice that
\begin{equation}
\sum_n^{\mathbb{Z}}\cdots\!\int_0^{L_x}\!\!\mathrm{d}x\cdots\!=
\sum_{n}^{\mathbb{Z}}
e^{i2\pi n(k_{2x}-k_{1x}-q_x)/N_x}\!\!
\int_{(n/N_x)L_x}^{\left(1+n/N_x\right)L_x}\!\!\!\!\mathrm{d}x\,
\exp\left\{
-\left[\frac{x}{l_B}
-\frac{2\pi l_B}{L_y}
\left(k_y+\frac{1}{2}\left(q_y-i\frac{L_y}{L_x}q_x\right)
\right)\right]^2
\right\}.
\end{equation}
Each interval $[\frac{m}{N_x}L_x,\frac{m+1}{N_x}L_x)$ is covered by the 
integral for $N_x$ times, and during the $N_x$ times, the exponential 
prefactor runs through all the $N_x$ values of $e^{i2\pi t(k_{2x}-k_{1x}-q_x)}$ 
for $t\in[0~..~N_x)$. In formula, we have
\begin{align}
\sum_n^{\mathbb{Z}}\cdots\int\mathrm{d}x\cdots
&=\sum_{n}^{N_x}
e^{i2\pi n(k_{2x}-k_{1x}-q_x)/N_x}\
\int_{-\infty}^{\infty}\mathrm{d}x\,
\exp\left\{
-\left[\frac{x}{l_B}
-\frac{2\pi l_B}{L_y}
\left(k_y+\frac{1}{2}\left(q_y-i\frac{L_y}{L_x}q_x\right)
\right)\right]^2
\right\}\\
&=\sqrt{\pi}N_xl_B\,\delta_{k_{2x},\;k_{1x}+q_x}^{\text{mod }N_x}.
\end{align}
The ``mod $N_x$'' does not lead to any problem, since $|\mathbf{k}\rangle$ is 
periodic in $k_x$.
Finally, we arrive at Eq.~\eqref{eq:density-Bloch}:
\begin{equation*}
\rho_\mathbf{q}
=e^{-\mathbf{q}^2l_B^2/4}
\sum_{\mathbf{k}}^{\mathrm{BZ}}
e^{-i2\pi q_x(k_y+q_y/2)/N_\phi}\,
|\mathbf{k}\rangle\langle{\mathbf{k}+\mathbf{q}}|.
\end{equation*}
\end{widetext}

\section{Pseudopotential Hamiltonian Reorganized}
\label{sec:sum-qx}
In this Appendix we prove Eq.~\eqref{eq:H-resummed}, the reorganized 
expression for the pseudopotential Hamiltonian in Eq.~\eqref{eq:H-Wannier} suitable 
for truncation.

Starting from Eq.~\eqref{eq:H-Wannier}, we first isolate the $q_x$ dependence,
\begin{multline}\label{eq:H-qx-isolated}
H=\sum_{q_y}^{\mathbb{Z}}
e^{-\frac{\pi}{N_\phi}\frac{L_x}{L_y}q_y^2}
\sum_{j_1}^M\!{}'\sum_{j_2}^M\!{}'
G_V(j_1-j_2,q_y)\\
\sum_{s_1}^{S_{j_1}}
\sum_{s_2}^{S_{j_2}}
\psi^\dagger_{j_1,s_1}
\psi^\dagger_{j_2,s_2}
\psi_{j_2-q_yd,s_2}
\psi_{j_1+q_yd,s_1},
\end{multline}
where $G_V(j_1-j_2,q_y)$ is defined by
\begin{multline}\label{eq:qx-sum}
G_V(k,q_y)=\\
\sum_{q_x}^{\mathbb{Z}}
\frac{V_\mathbf{q}}{2L_xL_y}
\exp\left[
-\frac{\pi}{N_\phi}\frac{L_y}{L_x}q_x^2
-i2\pi q_x\frac{k+q_yd}{M}
\right].
\end{multline}
Through a Poisson resummation, we can easily prove the general formula
\begin{equation}
\sum_{q_x}^\mathbb{Z}e^{-\pi A q_x^2-i2\pi q_x\xi}
=\frac{1}{\sqrt{A}}\sum_n^\mathbb{Z}e^{-\pi(n-\xi)^2/A}.
\end{equation}
Setting $A=L_y/(N_\phi L_x)$, $\xi=(k + q_yd)/M$, and defining
\begin{equation*}
\beta=\frac{1}{d^2}\frac{\pi}{N_\phi}\frac{L_x}{L_y},
\end{equation*}
we get
\begin{multline}\label{eq:qx-summed-basic}
\frac{1}{N_\phi}\sum_{q_x}^{\mathbb{Z}}
\exp\left[
-\frac{\pi}{N_\phi}\frac{L_y}{L_x}q_x^2
-i2\pi q_x\frac{k+q_yd}{M}
\right]\\
=\sqrt{\frac{L_x}{N_\phi L_y}}
\sum_n^\mathbb{Z}
e^{-\beta(k+q_yd-nM)^2}.
\end{multline}
To handle $G_V(k,q_y)$ in Eq.~\eqref{eq:qx-sum},
we need to be able to insert powers of $q_x^2$ into the $q_x$ sum.
This can be achieved by taking partial derivative with respect to 
$\frac{\pi L_y}{N_\phi L_x}=\beta d^2$ on 
Eq.~\eqref{eq:qx-summed-basic}.
For the simple case of
$V_\mathbf{q}=4\pi l_B^2 [V_0+V_1\cdot (1-\mathbf{q}^2l_B^2)]$
as in Eq.~\eqref{eq:V0-V1},
we find
\begin{multline}
G_V(k,q_y)= \sqrt{\frac{L_x}{N_\phi L_y}}
\sum_n^{\mathbb{Z}}
e^{-\beta(k+q_yd-nM)^2}\\
\left\{
V_0 + 2\beta V_1\left[(k+q_yd-nM)^2-(q_yd)^2\right]
\right\}.
\end{multline}
Plugging this back to Eq.~\eqref{eq:H-qx-isolated}, we get
\begin{multline}
H=C
\sum_{q_y}^{\mathbb{Z}}
e^{-\beta(q_yd)^2}
\sum_{j_1}^M\!{}'\sum_{j_2}^M\!{}'
\sum_n^{\mathbb{Z}}
e^{-\beta(j_1-j_2+q_yd-nM)^2}\\
\left\{
V_0 + 2\beta V_1\left[(j_1-j_2+q_yd-nM)^2-(q_yd)^2\right]
\right\}\\
\sum_{s_1}^{S_{j_1}}
\sum_{s_2}^{S_{j_2}}
\psi^\dagger_{j_1,s_1}
\psi^\dagger_{j_2,s_2}
\psi_{j_2-q_yd,s_2}
\psi_{j_1+q_yd,s_1},
\end{multline}
where $C=\sqrt{L_x/(N_\phi L_y)}$ is an inconsequential overall factor.

At last, recall from Eq.~\eqref{eq:j-s-range} that the range of summations 
over $j_1$ and $j_2$ each contain an arbitrary shift.
We can keep the outer sum over $j_1$ general, while make a convenient choice
for the inner sum over $j_2$.
We define $\Delta=j_2-j_1$ and rewrite the above equation as
\begin{multline}
H=C
\sum_{q_y}^{\mathbb{Z}}
e^{-\beta(q_yd)^2}
\sum_{j}^M\!{}'
\sum_{\Delta}
\sum_n^{\mathbb{Z}}
e^{-\beta(\Delta-q_yd+nM)^2}\\
\left\{
V_0 + 2\beta V_1\left[(\Delta-q_yd+nM)^2-(q_yd)^2\right]
\right\}\\
\sum_{s}^{S_{j}}
\sum_{s'}^{S_{j+\Delta}}
\psi^\dagger_{j,s}
\psi^\dagger_{j+\Delta,s'}
\psi_{j+\Delta-q_yd,s'}
\psi_{j+q_yd,s},
\end{multline}
where $\Delta$ is summed over an interval of length $M$ centered around $q_yd$,
\begin{equation}
\Delta\in\big[\,q_yd-\lfloor M/2\rfloor~..~q_yd-\lfloor M/2\rfloor+M\,\big).
\end{equation}
We make this special choice for the $\Delta$ sum
to facilitate later truncations in the thin-torus limit $\beta\gg 1$.
This proves Eq.~\eqref{eq:H-resummed}.

\end{document}